\documentclass[pdflatex,sn-basic]{sn-jnl}

\usepackage{graphicx}
\usepackage{lmodern}
\usepackage{amsmath,amssymb,amsfonts}
\usepackage{amsthm}
\usepackage{manyfoot}
\usepackage{mathtools}
\usepackage{booktabs}
\usepackage{array}
\usepackage{tikz}
\usetikzlibrary{arrows.meta}
\tikzset{
  sp/.style={circle,draw,thick,minimum size=9mm,inner sep=0pt,font=\small},
  ed/.style={-{Stealth[length=2.2mm]},thick},
}
\usepackage[dvipsnames]{xcolor}
\usepackage{enumitem}
\usepackage[title]{appendix}

\theoremstyle{plain}
\newtheorem{lemma}{Lemma}
\newtheorem{theorem}{Theorem}
\newtheorem{proposition}{Proposition}
\theoremstyle{remark}

\newcommand{\T}{\mathsf T}
\newcommand{\e}{\mathrm{e}}

\setcitestyle{aysep={},yysep={,},citesep={;}}
\hypersetup{bookmarksdepth=subsection}
\begin{document}


\title[Noise-induced Extinction in Hierarchical Food Webs]{Noise-induced Extinction in\hb Hierarchical Food Webs}

\author*[1]{\fnm{Varun} \sur{Nevash}}\email{varnvsh@engineering.upenn.edu }
\author[2]{\fnm{Eric} \sur{Forgoston}}
\author[1]{\fnm{M. Ani} \sur{Hsieh}}

\affil[1]{\orgdiv{Department of Mechanical Engineering and Applied Mechanics}, \orgname{University of Pennsylvania}, \orgaddress{\city{Philadelphia}, \state{PA} \postcode{19104}, \country{USA}}}
\affil[2]{\orgdiv{School of Computing}, \orgname{Montclair State University}, \orgaddress{\city{Montclair}, \state{NJ} \postcode{07043}, \country{USA}}}

\abstract{Every finite population is eventually driven to extinction by demographic noise, even when the deterministic dynamics have a stable coexistence state. We develop a large-deviation theory of this process for hierarchical food webs of arbitrary size. In a community of $N$ species, the basal species $X_1$ reproduces intrinsically and every consumer may prey on all species below it, so that omnivory is built into the interaction graph. For unit conversion efficiency and self-limitation confined to the basal species, the mean-field dynamics have at most $2^{N-1}+1$ feasible fixed points. We prove that the sign of each passive eigenvalue equals the sign of the extinct species' density in the enlarged community, that every partial community containing $X_1$ is a saddle with one unstable direction for each extinct consumer, and that the coexistence state is the only stable fixed point. A WKB treatment of the master equation gives a $2N$-dimensional Hamiltonian whose fluctuational fixed points are obtained in closed form. The resulting extinction network has $3^{N-1}$ fluctuation-driven transitions, and its coarse-grained dynamics yield an exact recursion for every mean extinction time. We analyse the three-species omnivory module in full, including a minimal set of existence conditions, global stability of coexistence, numerical instanton actions for all nine transitions and the most probable extinction route, which we confirm with stochastic simulations. Finally, we show how deleting a single trophic link reshapes the extinction network.}



\maketitle

\section{Introduction}

Demographic noise, the randomness of individual births, deaths and predation events, drives every isolated population to extinction, even when the deterministic dynamics possess a stable coexistence state. For small populations, this is a central concern of conservation biology, where demographic stochasticity sets a minimum viable population size \citep{Lande1993,Ovaskainen2010}. For large populations, extinction is a rare event. The community fluctuates about its coexistence state for a very long time before an unusually large fluctuation causes a species to go extinct. The mean time to extinction then grows exponentially with the characteristic population size $\Omega$, so direct stochastic simulation becomes prohibitively expensive and analytical tools are needed.

The theory of rare events has a long history. \citet{Kramers1940} treated the escape of a Brownian particle over a potential barrier and showed that the escape rate is exponentially small in the barrier height. Population models differ in two important respects. The state is discrete, and the noise is intrinsic, arising from the same reactions that drive the mean dynamics. \citet{Kubo1973} introduced an eikonal (WKB) ansatz for the master equation of such systems, while \cite{freidlin1984} established the rigorous large-deviation theory of the quasipotential. Later, \citet{Dykman1994} showed that large fluctuations follow optimal paths governed by an effective Hamiltonian.

\citet{Elgart2004} developed this Hamiltonian formulation for reaction systems. Working directly with the master equation in this way is essential. Although the diffusion (Fokker--Planck) approximation of the master equation captures typical fluctuations about the equilibrium, \citet{Doering2005} showed it generally predicts the wrong exponential scaling for rare extinction events. The WKB theory of extinction was subsequently refined through spectral and real-space formulations that resolve the discrete boundary layer near extinction and determine the pre-exponential factor of the mean extinction time \citep{AssafMeerson2006,Kessler2007,AssafMeerson2010}. Detailed reviews are provided by \citet{AssafMeerson2017} and \citet{ForgostonMoore2018}.

To illustrate how this framework works in its simplest analytically tractable setting, consider a population $X$ with birth rate $bX$, death rate $dX$ and competition rate $aX(X-1)/\Omega$, so that the density $x=X/\Omega$ obeys the logistic equation $\dot x=(b-d)x-ax^2$ with stable state $x^*=(b-d)/a$. Writing the quasi-stationary distribution as $P(x)\asymp\exp[-\Omega S(x)]$ turns the master equation, at leading order in $1/\Omega$, into the Hamilton--Jacobi equation $H(x,\partial_xS)=0$ with
\begin{equation*}
H(x,p) = b\,x\,(\e^{p}-1) + (d\,x + a\,x^2)(\e^{-p}-1).
\end{equation*}
Besides the deterministic line $p=0$, the zero-energy condition is satisfied on the fluctuational line $p(x)=\ln[(d+ax)/b]$, which connects the stable state $x^*$ to the extinction point. The action accumulated along it gives the mean extinction time
\begin{equation*}
\tau \asymp \e^{\Omega S}, \qquad S = \int_0^{x^*} \ln\frac{b}{d+a\,x}\,\mathrm dx = \frac{1}{a}\Big(b-d+d\ln\frac{d}{b}\Big).
\end{equation*}
For one-dimensional birth-death processes, therefore, the most likely path to extinction and its action are available in closed form.

In two or more dimensions this is no longer so. The optimal path to extinction, or instanton, is a heteroclinic orbit of Hamilton's equations in a phase space of twice the dimension, which connects the metastable equilibrium on the deterministic manifold $\mathbf p=0$ to the fluctuational extinct state. In general, the optimal path must be found numerically using shooting methods along unstable manifolds \citep{MeersonSasorov2009}, the minimum action method \citep{weinan2004} and its geometric arclength formulation gMAM \citep{heymann2008}, or the iterative action minimization method IAMM \citep{lindley2013}. 

With these numerical schemes, optimal extinction paths have been found for a variety of two- and higher-dimensional problems in population biology and epidemiology. In two dimensions, researchers have considered disease extinction under vaccination \citep{Dykman2008}, seasonal forcing in epidemic models \citep{billings2018}, and  Lotka--Volterra predator--prey dynamics \citep{ParkerKamenev2009}. In addition, \citet{GottesmanMeerson2012} considered a two-species predator-prey system and showed that extinction can occur along different routes, whose likelihoods are set by competing instanton actions. Therefore, the dominant route can switch as parameters change. Beyond two dimensions, optimal paths have been computed in higher-dimensional epidemic models \citep{bauver2016,nieddu2017}. In contrast, complex ecological communities with realistic trophic structure have received far less attention.

Natural food webs are rarely simple chains. Omnivory, in which a consumer feeds on more than one trophic level, is widespread \citep{PimmLawton1978,Polis1989}, and its simplest three-species form, intraguild predation, is a recognised community module with its own coexistence theory \citep{HoltPolis1997}. Omnivory alters both the stability of food webs \citep{McCann1997} and their deterministic fixed-point structure \citep{Hsu2015}, and it multiplies the ways in which species can become extinct.

Extending the instanton theory beyond two species raises several complications. First, extinction is no longer a single event. Species may go extinct one at a time or in multi-species cascades, and each partial extinction leaves a smaller community that is itself metastable. The natural object of study is therefore an extinction network, whose nodes are the feasible sub-communities and whose edges are the fluctuation-driven transitions. Second, each partial community must be characterised deterministically, since only states that are feasible and metastable can act as nodes.  Third, because the size of this network grows rapidly with the number of species, its nodes and edges must be classified systematically.  Fourth, every transition requires an instanton in a $2N$-dimensional phase space, whose end point, the fluctuational fixed point, generally lies off the deterministic manifold. Finally, the rates of individual transitions must be combined into mean extinction times for the whole community.

In this work we resolve these complications for a hierarchical food web of arbitrary size $N$, in which the basal species $X_1$ reproduces intrinsically and each consumer may prey on every species below it. For unit trophic conversion efficiency and self-limitation confined to the basal species, the interaction matrix is antisymmetric apart from a single diagonal term. This structure allows us to classify all feasible sub-communities and their stability in general, to obtain every fluctuational fixed point in closed form, and to show that the coarse-grained dynamics on the extinction network are governed by a triangular rate equation with an exact recursion for the mean extinction times. We then analyse the three-species omnivory module, the smallest food web containing omnivory. We compute the instanton actions for all nine transitions of its extinction network numerically, identify the most probable extinction route, and confirm the predictions with stochastic simulations.

Section~\ref{sec:general} introduces the hierarchical food web model and develops the general theory for $N$ species. Section~\ref{sec:three} specialises the theory to $N=3$, where the fixed points and their stability can be written in closed form, presents the numerical instanton actions and stochastic simulations, and analyses the reduced topologies. Section~\ref{sec:conclusions} summarises the results and discusses future work. Appendix~\ref{app:graphs} places the model within the combinatorial space of food-web graphs, Appendix~\ref{app:proofs} contains the proofs, and Appendix~\ref{app:stability} collects the stability criteria used in the three species case.

\section{Hierarchical food web of \texorpdfstring{$N$}{N} species}\label{sec:general}
\subsection{Model}\label{sec:model}

We represent a food web as a directed graph in which each vertex is a species, and each edge is a trophic interaction, oriented from prey to predator. The graphs are connected, so that they describe a single community, and contain no reciprocal predation (two different species preying on each other). When the species are strictly ranked so that every predator outranks its prey, the graph is acyclic, and we call such a web hierarchical. Relaxing the ranking admits directed cycles and greatly enlarges the number of possible topologies (Appendix~\ref{app:graphs}).

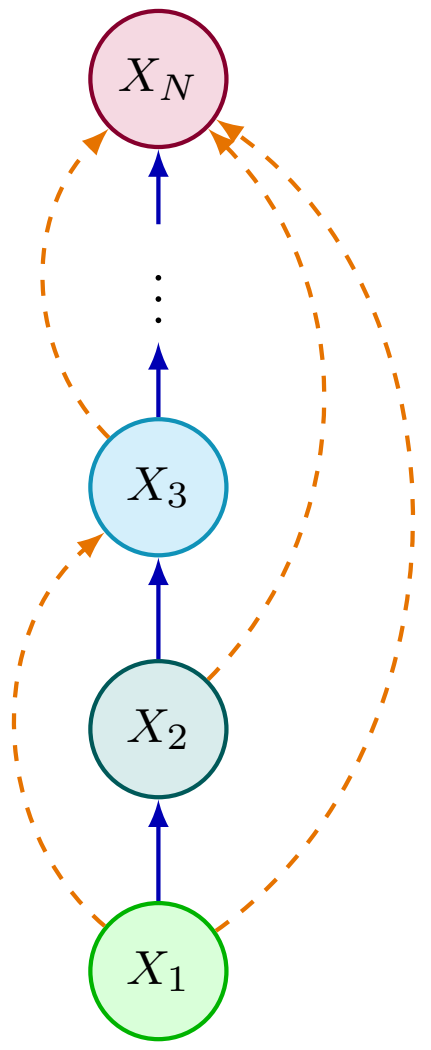
\begin{figure}[ht]
\centering
\begin{tikzpicture}[
  sp/.style={circle,draw=#1!70!black,fill=#1!15,thick,minimum size=9mm,inner sep=0pt,font=\small},
  ed/.style={-latex,thick,blue!70!black},
  edo/.style={-latex,thick,dashed,orange!90!black}]
\node[sp=green] (1) at (0,0) {$X_1$};
\node[sp=teal] (2) at (0,1.6) {$X_2$};
\node[sp=cyan] (3) at (0,3.2) {$X_3$};
\node (d) at (0,4.55) {$\vdots$};
\node[sp=purple] (N) at (0,5.9) {$X_N$};
\draw[ed] (1) -- (2);
\draw[ed] (2) -- (3);
\draw[ed] (3) -- (d);
\draw[ed] (d) -- (N);
\draw[edo] (1) to[bend left=50] (3);
\draw[edo] (1) to[bend right=55] (N);
\draw[edo] (2) to[bend right=45] (N);
\draw[edo] (3) to[bend left=45] (N);
\end{tikzpicture}
\caption{The complete trophic hierarchy for $N$ species: each consumer $X_j$ preys on every $X_i$ with $i<j$. Arrows point from prey to predator; links between adjacent levels are solid and omnivorous links, which skip a level, are dashed. For $N=3$ this is the omnivory (intraguild predation) module studied in Section~\ref{sec:three}}
\label{fig:hierarchy}
\end{figure}

We consider $N$ interacting species $X_1,\dots,X_N$ occupying successive levels of a strict trophic hierarchy $X_N\succ\dots\succ X_2\succ X_1$. The symbol $X_k$ denotes both the $k$-th species and its population (number of individuals). The populations are assumed to be well mixed, so the state of the system is the vector $\mathbf X=(X_1,\dots,X_N)$. The basal species $X_1$ grows on an external resource that is not modelled explicitly and preys on no other species. Every other species $X_j$, $j\ge2$, is a consumer and may prey on any species $X_i$ ranked below it, $i<j$. Any connected subset of the $N(N-1)/2$ allowed links gives a distinct hierarchical food web, of which there are A001187($N$)~$=1,1,4,38,728,\dots$ for $N=1,2,3,\dots$ (Appendix~\ref{app:graphs}). The theory below is developed for the complete hierarchy, in which every link is present (Fig.~\ref{fig:hierarchy}); for $N\ge3$ it contains omnivorous links, in which a consumer feeds on more than one trophic level. This is only one of the possible topologies, but is the densest, and every other hierarchical web is obtained from it by deleting links, i.e.\ by setting the corresponding predation rates to zero. For $N=3$ the complete web is treated in Section~\ref{sec:three}, and the three webs obtained by deleting a single link are analysed in Section~\ref{sec:reducedtop}.

The system size $\Omega\gg1$ sets the typical population scale,  and all asymptotic statements are to leading order in $1/\Omega$. The elementary processes are listed in Table~\ref{tab:processes}. Each species is born at per-capita rate $b_k$, dies or emigrates at per-capita rate $d_k$, and experiences intraspecific competition at rate $a_{kk}/\Omega$ per pair. Predation of $X_i$ by $X_j$ ($i<j$) occurs at rate $a_{ij}/\Omega$ per pair; with probability $e_{ij}$ the consumed prey is converted into a new predator, and otherwise it is simply removed. This formulation keeps every stoichiometric coefficient integer while reproducing the usual trophic efficiency at the mean-field level.

\begin{table}[ht]
\centering
\caption{Elementary processes of the $N$-species food web ($1\le k\le N$, $1\le i<j\le N$)}
\label{tab:processes}
\begin{tabular}{lll}
\toprule
Process & Description & Transition rate \\
\midrule
$X_k\to 2X_k$ & intrinsic reproduction & $b_kX_k$ \\
$X_k\to\varnothing$ & death or emigration & $d_kX_k$ \\
$2X_k\to X_k$ & intraspecific competition & $(a_{kk}/\Omega)\,X_k(X_k-1)$ \\
$X_i+X_j\to X_j$ & predation without conversion & $(1-e_{ij})(a_{ij}/\Omega)\,X_iX_j$ \\
$X_i+X_j\to 2X_j$ & predation with conversion & $e_{ij}(a_{ij}/\Omega)\,X_iX_j$ \\
\bottomrule
\end{tabular}
\end{table}

Let $x_k=X_k/\Omega$ be the density of $X_k$ and $\mathbf x=(x_1,\dots,x_N)$. Taking $\Omega\to\infty$, the densities obey the mean-field equations
\begin{equation}\label{eq1}
\dot x_k = (b_k-d_k)\,x_k - a_{kk}\,x_k^2 + \sum_{i<k} e_{ik}\,a_{ik}\,x_i x_k - \sum_{j>k} a_{kj}\,x_k x_j, \qquad k = 1,\dots,N.
\end{equation}
The full model \eqref{eq1} contains $N(N+2)$ parameters: $b_k$, $d_k$ and $a_{kk}$ for each species, and $a_{ij}$ and $e_{ij}$ for each link.

\subsection{Reduced model}\label{sec:reduced}

To obtain a model whose fixed points can be classified for every $N$, we impose three simplifying assumptions.
\begin{enumerate}
\item \textbf{Consumers reproduce only through predation:} $b_k=0$ for $k\ge2$, so that their net intrinsic growth rate is $-d_k<0$. We write $B_1\equiv b_1-d_1$ for the net intrinsic growth rate of the basal species.
\item \textbf{Self-limitation acts only at the base:} $a_{kk}=0$ for $k\ge2$, on the grounds that consumer populations are regulated mainly by their prey and predators.
\item \textbf{Unit conversion efficiency:} $e_{ij}=1$ for all links, so that every predation event is the conversion $X_i+X_j\to2X_j$.
\end{enumerate}
Assumption 3 is a parametric simplification rather than an ecological claim. Its role is structural: it makes the interaction matrix antisymmetric apart from the basal self-limitation term, a property on which the stability results of Section~\ref{sec:stability} rest. Under these assumptions \eqref{eq1} becomes
\begin{equation}\label{eq2}
\dot x_1 = x_1\Big(B_1 - a_{11}x_1 - \sum_{j=2}^{N} a_{1j}x_j\Big),
\end{equation}
\begin{equation}\label{eq3}
\dot x_k = x_k\Big(-d_k + \sum_{i<k} a_{ik}x_i - \sum_{j>k} a_{kj}x_j\Big), \qquad k = 2,\dots,N.
\end{equation}
In compact form,
\begin{equation}\label{eq4}
\dot x_k = x_k\,g_k(\mathbf x), \qquad \mathbf g(\mathbf x) = \mathbf r + \mathbf K\mathbf x, \qquad \mathbf K = \boldsymbol\Lambda - a_{11}\,\mathbf e_1\mathbf e_1^{\T},
\end{equation}
where $\mathbf r=(B_1,-d_2,\dots,-d_N)$, $\mathbf e_1$ is the first unit vector, and $\boldsymbol\Lambda$ is the antisymmetric matrix with $\Lambda_{kj}=a_{jk}$ for $j<k$ and $\Lambda_{kj}=-a_{kj}$ for $j>k$. The reduced model has $(N^2+N+4)/2$ parameters: $b_1,d_1,\dots,d_N$, $a_{11}$ and the $N(N-1)/2$ predation rates. The mean-field dynamics depend on $b_1$ and $d_1$ only through $B_1$, but the stochastic dynamics depend on them separately. 

General Lotka--Volterra systems with three or more species can have limit cycles and chaotic attractors \citep{Arneodo1980,Arneodo1982}, for which extinction would have to be treated from a non-stationary attractor. The assumptions above exclude this for $N=3$, since the Volterra function of Proposition~\ref{prop:lyapunov} forces every interior trajectory to converge to a fixed point (Proposition~\ref{prop:global}). For general $N$ the same function rules out unbounded growth and deterministic extinction, but not non-stationary attractors on the set $\{x_1=x_1^*\}$. Throughout, we consider parameters for which the deterministic attractor is a fixed point, so that every metastable state of the stochastic dynamics is a fixed point and every extinction is a transition between fixed points.
\subsection{Feasible fixed points}\label{sec:fixedpoints}

A fixed point of \eqref{eq2} and \eqref{eq3} is characterised by its {active set} $A\subseteq\{1,\dots,N\}$, the set of species with $x_k>0$. We say that species $X_k$ persists at the fixed point if $k\in A$ and is extinct if $k\notin A$, i.e.\ if $x_k=0$. Since only the basal species reproduces intrinsically, no consumer can persist without $X_1$. Indeed, if $A$ is non-empty and $1\notin A$, let $k$ be the smallest index in $A$. Every prey of $X_k$ is then extinct, so its per-capita growth rate $g_k=-d_k-\sum_{j>k}a_{kj}x_j$ is strictly negative and $x_k$ cannot remain at a positive stationary value. The admissible active sets are therefore the empty set and the $2^{N-1}$ subsets that contain $1$, which gives
\begin{equation}\label{eq5}
\#\{\text{feasible fixed points}\} \;\le\; 2^{N-1} + 1.
\end{equation}
The bound is attained only when every admissible active set yields a fixed point at which all species in the active set persist, i.e.\ have strictly positive density. Whether this holds depends on the parameters, through the positivity conditions derived below.

We write $M_A$ for the fixed point with active set $A$, abbreviating $M_0\equiv M_\varnothing$, $M_1\equiv M_{\{1\}}$, $M_{1j}\equiv M_{\{1,j\}}$, and so on. Two of these states play a special role: the coexistence state $M_{12\cdots N}$, at which all $N$ species persist, and the state of total extinction $M_0$, at which every species is extinct. Every other feasible fixed point describes a partial community, in which the basal species persists together with some, but not all, of the consumers. Equivalently, since $1$ belongs to every non-empty active set, the fixed points with $n\ge1$ persisting species may be enumerated as $M_{n,i}$, $i=1,\dots,\binom{N-1}{n-1}$, by ordering their active sets lexicographically. For example, for $N=4$ the $2^{3}+1=9$ admissible fixed points are
\begin{equation}\label{eq:N4labels}
\begin{array}{lll}
n=0: & M_0, & \\
n=1: & M_1 = M_{1,1}, & \\
n=2: & M_{12}=M_{2,1},\quad M_{13}=M_{2,2},\quad M_{14}=M_{2,3}, & \\
n=3: & M_{123}=M_{3,1},\quad M_{124}=M_{3,2},\quad M_{134}=M_{3,3}, & \\
n=4: & M_{1234}=M_{4,1}, &
\end{array}
\end{equation}
so that, with the basal species always present, there are three distinct two-species communities and three distinct three-species communities, and $M_{1234}$ is the coexistence state.

Let $A$ have $n=|A|$ elements, listed in increasing order as $A=\{s_1,s_2,\dots,s_n\}$ with $s_1=1<s_2<\dots<s_n$, so that $s_i$ is the index of the $i$-th persisting species and $s_n$ is the highest-ranked one. Write $\mathbf x_A=(x_{s_1},\dots,x_{s_n})^{\T}$ for the densities of the persisting species. The stationarity conditions $g_{s_i}=0$, $i=1,\dots,n$, then form the linear system $\mathbf L_A\mathbf x_A=\mathbf c_A$, with
\begin{equation}\label{eq6}
\mathbf L_A = \begin{pmatrix} a_{11} & a_{1s_2} & a_{1s_3} & \cdots & a_{1s_n} \\ a_{1s_2} & 0 & -a_{s_2s_3} & \cdots & -a_{s_2s_n} \\ a_{1s_3} & a_{s_2s_3} & 0 & \cdots & -a_{s_3s_n} \\ \vdots & \vdots & \vdots & \ddots & \vdots \\ a_{1s_n} & a_{s_2s_n} & a_{s_3s_n} & \cdots & 0 \end{pmatrix}, \qquad \mathbf c_A = \begin{pmatrix} B_1 \\ d_{s_2} \\ \vdots \\ d_{s_n} \end{pmatrix}.
\end{equation}

The first row is the basal equation, and the following rows $\ell\ge2$ are the consumer equations of $X_{s_\ell}$. Write $\mathbf L_A$ in block form, with the scalar $a_{11}$, the vector $\mathbf u_A=(a_{1s_2},\dots,a_{1s_n})$ and the  $(n-1)\times(n-1)$ antisymmetric lower-right block $\boldsymbol\Lambda'_A$. This form determines the sign of the determinant.

\begin{lemma}[Sign of the determinant]\label{lem:det}
For every admissible $A$ with $|A|=n$,
\begin{equation}\label{eq7}
\det \mathbf L_A = \begin{cases} a_{11}\,\operatorname{Pf}(\boldsymbol\Lambda'_A)^2, & n \text{ odd},\\[2pt] -\big(\mathbf u_A^{\T}\mathbf w_A\big)^2, & n \text{ even}, \end{cases} \qquad\text{so that}\qquad (-1)^{n+1}\det \mathbf L_A \ge 0,
\end{equation}
where $\operatorname{Pf}$ denotes the Pfaffian and, for $n$ even, $\mathbf w_A$ spans the kernel of $\boldsymbol\Lambda'_A$ with $\operatorname{adj}(\boldsymbol\Lambda'_A)=\mathbf w_A\mathbf w_A^{\T}$.
\end{lemma}

The proof is given in Appendix~\ref{app:proofs}. The first three cases are $\det \mathbf L_{\{1\}}=a_{11}$, $\det \mathbf L_{\{1,j\}}=-a_{1j}^2$ and $\det \mathbf L_{\{1,j,k\}}=a_{11}a_{jk}^2$. In particular, a three-species state $M_{1jk}$ cannot exist unless $X_j$ and $X_k$ interact directly ($a_{jk}\neq0$). We assume throughout that the parameters are generic, so that every $\det \mathbf L_A$ is non-zero and the coordinates of $M_A$ follow from Cramer's rule, $x_s=\det \mathbf L_A^{(s)}/\det \mathbf L_A$, where $\mathbf L_A^{(s)}$ is $\mathbf L_A$ with the column of species $s$ replaced by $\mathbf c_A$.

The low-order fixed points are
\begin{equation}\label{eq8}
M_0 = (0,\dots,0), \qquad M_1:\ x_1 = \frac{B_1}{a_{11}}, \quad \text{feasible iff } B_1 > 0,
\end{equation}
\begin{equation}\label{eq9}
M_{1j}:\ x_1 = \frac{d_j}{a_{1j}}, \quad x_j = \frac{B_1 a_{1j} - a_{11} d_j}{a_{1j}^2}, \quad \text{feasible iff } B_1 a_{1j} > a_{11} d_j,
\end{equation}
and, for $1<j<k$,
\begin{equation}\label{eq10}
M_{1jk}:\quad \begin{aligned} x_1 &= \frac{B_1 a_{jk} - a_{1j} d_k + a_{1k} d_j}{a_{11} a_{jk}}, \\ x_j &= \frac{a_{11} a_{jk} d_k - B_1 a_{jk} a_{1k} + a_{1j} a_{1k} d_k - a_{1k}^2 d_j}{a_{11} a_{jk}^2}, \\ x_k &= \frac{B_1 a_{1j} a_{jk} - a_{11} a_{jk} d_j - a_{1j}^2 d_k + a_{1j} a_{1k} d_j}{a_{11} a_{jk}^2}. \end{aligned}
\end{equation}

Feasibility of $M_A$ requires $|A|$ positivity conditions, one per active coordinate. Summed over all admissible active sets this gives $\sum_n n\binom{N-1}{n-1}=(N+1)\,2^{N-2}$ conditions, of which the $N-1$ conditions $x_1=d_j/a_{1j}>0$ hold automatically. These conditions are not independent: for $N=3$ the six non-trivial ones reduce to three (Proposition~\ref{prop:existence}, Section~\ref{sec:three-fp}). Identifying a minimal independent set for general $N$ is an open problem.

Unless stated otherwise, we assume that all $2^{N-1}+1$ fixed points in \eqref{eq5} are feasible. This is the richest case, and the one in which every extinction route is available.

\subsection{Linear stability}\label{sec:stability}

Differentiating \eqref{eq4}, the Jacobian of the reduced model is
\begin{equation}\label{eq11}
J_{kl}(\mathbf x) = \delta_{kl}\,g_k(\mathbf x) + x_k K_{kl}.
\end{equation}
At $M_A$ every extinct species $k\notin A$ has $x_k=0$, so the $k$-th row of $\mathbf J$ vanishes except for its diagonal entry. Listing the active species first, the Jacobian is block upper-triangular,
\begin{equation}\label{eq12}
\mathbf J_A = \begin{pmatrix} \mathbf J_A^{S} & \ast \\ 0 & \mathbf D_A \end{pmatrix}, \qquad \mathbf J_A^{S} = \mathbf X_A \mathbf K_A, \qquad \mathbf D_A = \operatorname{diag}\big(f_k^{(A)}\big)_{k\notin A}
\end{equation}
where $\mathbf X_A=\operatorname{diag}(x_s)$ and $\mathbf K_A$ is the restriction of $\mathbf K$ to the active species. Its spectrum is therefore the union of the spectrum of the active block $\mathbf J_A^S$ and the passive eigenvalues $f_k^{(A)}=g_k(M_A)$, the per-capita growth rates of rare entrants. For a consumer $k\ge2$,
\begin{equation}\label{eq13}
f_k^{(A)} = G_k^{(A)} - D_k^{(A)}, \qquad G_k^{(A)} = \sum_{i<k,\, i\in A} a_{ik}\,x_i, \qquad D_k^{(A)} = d_k + \sum_{j>k,\, j\in A} a_{kj}\,x_j,
\end{equation}
the difference between the rate at which an entrant gains from its prey and the rate at which it is lost to death and to its predators. The quantities $G_k$ and $D_k$ reappear as the arguments of the fluctuational momenta in Section~\ref{sec:fluct}.

Here an entrant is a species $X_k$, $k\notin A$, that is extinct at $M_A$. Since the model has no immigration, an extinct species cannot return, and the hyperplane $x_k=0$ is invariant under the dynamics. We therefore consider a hypothetical reintroduction of $X_k$ at a density so small that it does not affect the existing species. This is a mathematical device for probing the stability of $M_A$ in the direction transverse to the hyperplane, not a process that occurs in the dynamics. While rare, the density of the entrant evolves as $\dot x_k\simeq f_k^{(A)}x_k$, so $X_k$ establishes itself if $f_k^{(A)}>0$ and fails to do so if $f_k^{(A)}<0$. We call $f_k^{(A)}$ the passive eigenvalue of $X_k$ at $M_A$. 

\begin{lemma}[Establishment criterion]\label{lem:establishment}
Let $1\in A$, $k\notin A$ and $A'=A\cup\{k\}$, and let $x_k^{(A')}$ be the $k$-th coordinate of the solution of $\mathbf L_{A'}\mathbf x=\mathbf c_{A'}$, whether or not $M_{A'}$ is feasible. Then
\begin{equation}\label{eq14}
f_k^{(A)} = -\,\frac{\det \mathbf L_{A'}}{\det \mathbf L_{A}}\; x_k^{(A')},
\end{equation}
and the prefactor is strictly positive by Lemma~\ref{lem:det}. Hence $X_k$ can establish itself in the community $A$ if and only if its own density at the enlarged fixed point is positive.
\end{lemma}

Lemma~\ref{lem:establishment} concerns only the entrant's coordinate. In addition, the feasibility of $M_{A'}$ requires that the coordinates of the resident species remain positive.

\begin{lemma}[Active block]\label{lem:active}
For every feasible $M_A$ with $1\in A$, every eigenvalue $\lambda$ of $\mathbf J_A^S$ satisfies $\operatorname{Re}\lambda\le0$.
\end{lemma}

The proof uses only that $\mathbf X_A$ is positive definite and that $\mathbf K_A$ is antisymmetric apart from the single entry $-a_{11}$ in its top-left corner. Both lemmas are proved in Appendix~\ref{app:proofs}.

Lotka--Volterra systems with antisymmetric interactions have been studied extensively in the deterministic setting. The surviving species are characterised by a set of linear inequalities \citep{XuLiuLou2024}, and in conservative networks the Pfaffian of the interaction matrix decides between coexistence and extinction \citep{Knebel2013}. An antisymmetric matrix of even order is generically invertible, so an even number of species cannot all coexist in such a network, and the surviving set contains an odd number of species \citep{Knebel2013}. The basal self-limitation term $-a_{11}\mathbf e_1\mathbf e_1^{\mathsf T}$ makes the present system dissipative, so that the coexistence state is a metastable attractor and extinction times grow exponentially with $\Omega$, rather than algebraically as in the conservative case.

\begin{proposition}[Lyapunov stability of coexistence]\label{prop:lyapunov}
Suppose the coexistence state $\mathbf x^*=M_{12\cdots N}$ is feasible. Then the Volterra function \citep{HofbauerSigmund1998}
\begin{equation}\label{eq15}
V(\mathbf x) = \sum_{k=1}^{N}\Big(x_k - x_k^{*} - x_k^{*}\ln\frac{x_k}{x_k^{*}}\Big)
\end{equation}
satisfies $\dot V = -a_{11}(x_1-x_1^{*})^2 \le 0$ in the interior of the positive orthant. Hence $\mathbf x^*$ is Lyapunov stable for every $N$, and it attracts every interior trajectory whenever the largest invariant subset of $\{x_1=x_1^*\}$ is $\mathbf x^*$ itself.
\end{proposition}

\begin{proof}
The derivative $\dot V$ is computed in Appendix~\ref{app:proofs}. Each term of $V$ equals $x_k^*\,h(x_k/x_k^*)$ with $h(y)=y-1-\ln y\ge0$, which vanishes only at $y=1$, so $V>0$ away from $\mathbf x^*$ and $V$ is a Lyapunov function. The attraction statement follows from LaSalle's invariance principle, and its condition is verified for $N=3$ in Section~\ref{sec:three-stab}.
\end{proof}

\begin{theorem}[Saddle classification]\label{thm:saddle}\hfill
\begin{enumerate}[leftmargin=*, topsep=0pt]
\item $M_0$ has eigenvalues $B_1>0$ and $-d_2,\dots,-d_N<0$. It is a saddle with one unstable direction.
\item Every $M_A$ with $1\in A\subsetneq\{1,\dots,N\}$ has $N-|A|$ strictly positive passive eigenvalues, and its remaining $|A|$ eigenvalues have non-positive real part. It is a saddle with $N-|A|$ unstable directions, one for each extinct consumer.
\item The coexistence state $M_{12\cdots N}$ has no extinct species, and all its eigenvalues have non-positive real part. It is the only feasible fixed point that is Lyapunov stable.
\end{enumerate}
\end{theorem}

\begin{proof}
Part 1 follows because $\mathbf J$ is diagonal at the origin. For part 2, the existence assumption makes $M_{A'}$ feasible for every $k\notin A$, so $x_k^{(A')}>0$ and Lemma~\ref{lem:establishment} gives $f_k^{(A)}>0$. Lemma~\ref{lem:active} controls the active block. Part 3 follows from Lemma~\ref{lem:active} with $A=\{1,\dots,N\}$ and Proposition~\ref{prop:lyapunov}.
\end{proof}

When the coexistence state is feasible, the deterministic dynamics therefore never lose a species, and every loss of species from this state is driven by demographic fluctuations.

\subsection{Master equation and WKB Hamiltonian}\label{sec:wkb}

Under the assumptions of Section~\ref{sec:reduced}, the probability $P(\mathbf X,t)$ of finding $\mathbf X=(X_1,\dots,X_N)$ at time $t$ obeys the master equation
\begin{equation}\label{eq16}
\begin{aligned} \partial_t P(\mathbf X) ={}& b_1\big[(X_1-1)P(\mathbf X-\mathbf e_1) - X_1P(\mathbf X)\big] \\ &+ \frac{a_{11}}{\Omega}\big[(X_1+1)X_1P(\mathbf X+\mathbf e_1) - X_1(X_1-1)P(\mathbf X)\big] \\ &+ \sum_{k=1}^{N} d_k\big[(X_k+1)P(\mathbf X+\mathbf e_k) - X_kP(\mathbf X)\big] \\ &+ \sum_{i<j}\frac{a_{ij}}{\Omega}\big[(X_i+1)(X_j-1)P(\mathbf X+\mathbf e_i-\mathbf e_j) - X_iX_jP(\mathbf X)\big], \end{aligned}
\end{equation}
where $\mathbf e_k$ is the $k$-th unit vector. All extinction rates can be shown to be exponentially small in $\Omega$, so on time scales that are short compared with their inverses the probability in the interior of the state space is quasi-stationary. Substituting the eikonal ansatz $P(\mathbf x,t) \asymp \exp[-\Omega S(\mathbf x,t)]$ and expanding to leading order in $1/\Omega$ gives the Hamilton-Jacobi equation $\partial_tS+H(\mathbf x,\partial S/\partial\mathbf x)=0$, whose quasi-stationary form is $H(\mathbf x,\nabla S)=0$, with the Hamiltonian
\begin{equation}\label{eq17}
\begin{aligned}
H(\mathbf x,\mathbf p) ={}& b_1x_1\big(\e^{p_1}-1\big) + \big(d_1x_1 + a_{11}x_1^2\big)\big(\e^{-p_1}-1\big) \\
&+ \sum_{k=2}^{N} d_kx_k\big(\e^{-p_k}-1\big) + \sum_{i<j} a_{ij}\,x_ix_j\big(\e^{p_j-p_i}-1\big).
\end{aligned}
\end{equation}
Each term corresponds to one elementary process $r$, so that \eqref{eq17} has the general form
\begin{equation}\label{eq:Hgeneral}
H(\mathbf x,\mathbf p)=\sum_r W_r(\mathbf x)\big(\e^{\boldsymbol\nu_r\cdot\mathbf p}-1\big),
\end{equation}
where $W_r(\mathbf x)$ is the rate of process $r$ per unit $\Omega$ and $\boldsymbol\nu_r$ is the change in the population vector $\mathbf X$. The momentum $p_k=\partial S/\partial x_k$ is conjugate to $x_k$, and Hamilton's equations $\dot x_k=\partial H/\partial p_k$, $\dot p_k=-\partial H/\partial x_k$ are given as
\begin{equation}\label{eq18}
\dot x_1 = b_1x_1\e^{p_1} - \big(d_1x_1 + a_{11}x_1^2\big)\e^{-p_1} - \sum_{j\ge2} a_{1j}\,x_1x_j\,\e^{p_j-p_1},
\end{equation}
\begin{equation}\label{eq19}
\dot x_k = -d_kx_k\e^{-p_k} + \sum_{i<k} a_{ik}\,x_ix_k\,\e^{p_k-p_i} - \sum_{j>k} a_{kj}\,x_kx_j\,\e^{p_j-p_k}, \qquad k\ge2,
\end{equation}
\begin{equation}\label{eq20}
\dot p_1 = b_1\big(1-\e^{p_1}\big) + \big(d_1 + 2a_{11}x_1\big)\big(1-\e^{-p_1}\big) + \sum_{j\ge2} a_{1j}\,x_j\big(1-\e^{p_j-p_1}\big),
\end{equation}
\begin{equation}\label{eq21}
\dot p_k = d_k\big(1-\e^{-p_k}\big) + \sum_{i<k} a_{ik}\,x_i\big(1-\e^{p_k-p_i}\big) + \sum_{j>k} a_{kj}\,x_j\big(1-\e^{p_j-p_k}\big), \qquad k\ge2.
\end{equation}

Four properties of this system are:
\begin{enumerate}
\item \textbf{Deterministic manifold.} The submanifold $\mathbf p=0$ is invariant, and on it \eqref{eq18} and \eqref{eq19} reduce to the mean-field equations \eqref{eq2} and \eqref{eq3}. Every $M_A$ is therefore a fixed point of the Hamiltonian flow with $\mathbf p=0$.
\item \textbf{Extinction hyperplanes.} Each hyperplane $x_k=0$ is invariant, because $\dot x_k$ is proportional to $x_k$.
\item \textbf{Phase-space dimension.} Linearising about $(M_A,\mathbf p=0)$ gives $\delta\dot{\mathbf x}=\mathbf J_A\,\delta\mathbf x+\mathbf Q\,\delta\mathbf p$ and $\delta\dot{\mathbf p}=-\mathbf J_A^{\T}\delta\mathbf p$, where $\mathbf Q=\sum_r W_r\,\boldsymbol\nu_r\boldsymbol\nu_r^{\T}$, evaluated at $M_A$ with $W_r$ and $\boldsymbol\nu_r$ as in \eqref{eq:Hgeneral}, is the diffusion matrix of the demographic noise. Since the momentum equation does not involve $\delta\mathbf x$, the linearised system is block triangular, and its $2N$ eigenvalues are $\pm\lambda$ for each eigenvalue $\lambda$ of $\mathbf J_A$. A hyperbolic $M_A$ therefore has $N$-dimensional stable and unstable manifolds in phase space, irrespective of the number of unstable directions in the mean-field dynamics.
\item \textbf{Instantons.} Typical fluctuations about a metastable state are of order $\sqrt\Omega$, whereas losing a species requires an excursion of order $\Omega$. To exponential accuracy such excursions follow an instanton, a zero-energy trajectory that leaves $M_A$ as $t\to-\infty$ and approaches a fluctuational fixed point (Section~\ref{sec:fluct}) as $t\to+\infty$, and
\begin{equation}\label{eq22}
r_{A\to B} \asymp \exp\big(-\Omega\,S_{A\to B}\big), \qquad S_{A\to B} = \int_{-\infty}^{\infty} \mathbf p\cdot\dot{\mathbf x}\;\mathrm dt,
\end{equation}
where the integral is taken along the instanton and $\asymp$ denotes equality to exponential accuracy, that is,
\begin{equation}\label{eq:asymp}
\lim_{\Omega\to\infty} \left(-\frac{1}{\Omega}\ln r_{A\to B}\right) = S_{A\to B}.
\end{equation}
The full rate carries a pre-exponential factor,
\begin{equation}
r_{A\to B} = K_{A\to B}\,\Omega^{\alpha}\,\e^{-\Omega S_{A\to B}}\big[1+O(\Omega^{-1})\big],
\end{equation}
where the constant $K_{A\to B}$ and the exponent $\alpha$ are independent of $\Omega$ and depend on the local behaviour of the dynamics near $M_A$ and $F_B$, which are determined by the next order of the WKB expansion. The instanton is a heteroclinic connection in the full $2N$-dimensional phase space. A species that is extinct at both ends of a transition has $x_k=0$ along the whole path, but its momentum $p_k$ need not vanish there. Every term of $H$ involving $X_k$ carries a factor $x_k$, so the momentum $p_k$ does not affect the action, but it must be retained for the instanton to end at the fluctuational point $F_B$.
\end{enumerate}

\subsection{Fluctuational fixed points}\label{sec:fluct}

Let $M_A$ be a feasible fixed point with $1\in A$, and set $p_i=0$ for every active species $i\in A$. Every term in \eqref{eq18}--\eqref{eq21} that couples an active species to an extinct one carries a factor $x_k=0$, so all $\dot x$ vanish, all active $\dot p_i$ vanish, and, for every extinct consumer $k\notin A$, equation \eqref{eq21} reduces to
\begin{equation}\label{eq23}
\dot p_k = D_k^{(A)}\big(1-\e^{-p_k}\big) + G_k^{(A)}\big(1-\e^{p_k}\big),
\end{equation}
with $G_k$ and $D_k$ the gain and loss rates of \eqref{eq13}. The extinct momenta therefore decouple from one another. Each has the deterministic root $p_k=0$ and the fluctuational root
\begin{equation}\label{eq24}
p_k^{*} = \ln\frac{D_k^{(A)}}{G_k^{(A)}}, \qquad \e^{p_k^{*}} = 1 - \frac{f_k^{(A)}}{G_k^{(A)}},
\end{equation}
so that $p_k^*<0$ exactly when $X_k$ could establish itself in the community $A$. Choosing, for each extinct consumer, either $0$ or $p_k^*$ gives the fluctuational fixed points $F_A^{(S)}$, labelled by the non-empty set $S\subseteq\{1,\dots,N\}\setminus A$ of species whose momenta are non-zero. Since $G_k\ge a_{1k}x_1>0$ whenever $1\in A$, all $2^{N-|A|}-1$ of them exist. They lie on the zero-energy surface, because every term of \eqref{eq17} vanishes there. At the origin the consumers have no prey, $G_k=0$, and their momenta have only the root $0$, while the basal momentum solves $b_1(1-\e^{p_1})+d_1(1-\e^{-p_1})=0$. The only fluctuational point at $M_0$ is
\begin{equation}\label{eq25}
F_0 = \Big(\mathbf 0;\ \ln\frac{d_1}{b_1},\,0,\dots,0\Big).
\end{equation}
Counting all of them,
\begin{equation}\label{eq26}
1 + \sum_{n=1}^{N}\binom{N-1}{n-1}\big(2^{N-n}-1\big) = 3^{N-1} - 2^{N-1} + 1.
\end{equation}

The instanton for a transition $M_A\to M_B$ with $1\in B$ terminates at some $F_B^{(S)}$ with $S\supseteq A\setminus B$, since the species lost in the transition must end on their fluctuational roots. Whether $S$ also contains species that were already extinct at $M_A$ is not fixed by this argument and must be decided by comparing the actions of the candidate instantons. A transition $M_A\to M_0$ is triggered by the loss of $X_1$ and terminates at $F_0$.

\subsection{Extinction network and mean extinction times}\label{sec:network}

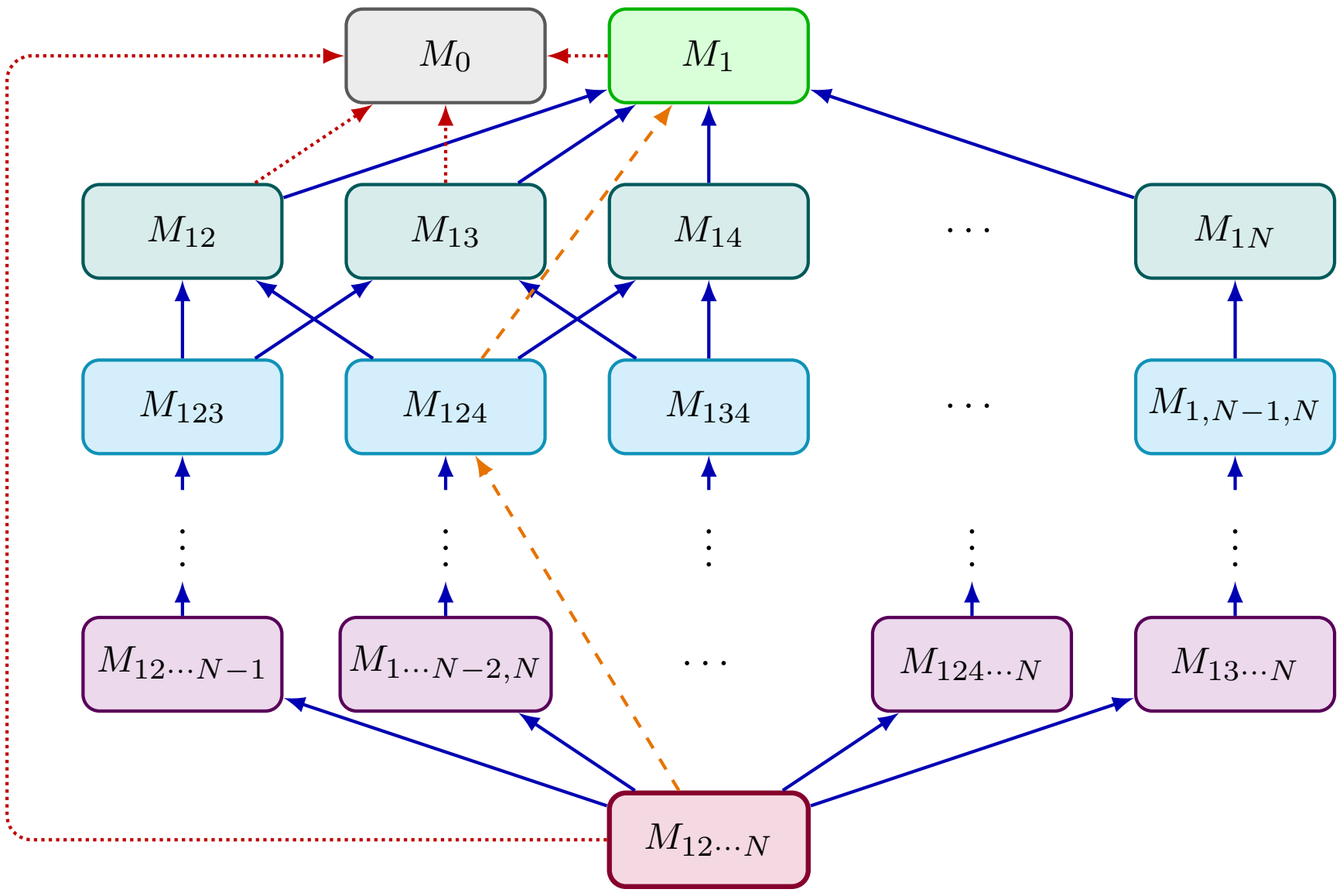
\begin{figure}[ht]
\centering
\begin{tikzpicture}[x=1cm,y=1cm,
  st/.style={rectangle,rounded corners=4pt,draw=#1!70!black,fill=#1!15,thick,minimum width=17mm,minimum height=8mm,inner sep=2pt,font=\small},
  ed/.style={-latex,thick,blue!70!black},
  edj/.style={-latex,thick,dashed,orange!90!black},
  ed0/.style={-latex,thick,densely dotted,red!75!black}]
\node[st=gray] (0) at (-2.25,6.2) {$M_0$};
\node[st=green] (1) at (0,6.2) {$M_1$};
\node[st=teal] (12) at (-4.5,4.7) {$M_{12}$};
\node[st=teal] (13) at (-2.25,4.7) {$M_{13}$};
\node[st=teal] (14) at (0,4.7) {$M_{14}$};
\node at (2.25,4.7) {$\cdots$};
\node[st=teal] (1N) at (4.5,4.7) {$M_{1N}$};
\node[st=cyan] (123) at (-4.5,3.2) {$M_{123}$};
\node[st=cyan] (124) at (-2.25,3.2) {$M_{124}$};
\node[st=cyan] (134) at (0,3.2) {$M_{134}$};
\node at (2.25,3.2) {$\cdots$};
\node[st=cyan] (1MN) at (4.5,3.2) {$M_{1,N-1,N}$};
\node (v1) at (-4.5,2.1) {$\vdots$};
\node (v2) at (-2.25,2.1) {$\vdots$};
\node (v3) at (0,2.1) {$\vdots$};
\node (v4) at (2.25,2.1) {$\vdots$};
\node (v5) at (4.5,2.1) {$\vdots$};
\node[st=violet] (a) at (-4.5,1.0) {$M_{12\cdots N-1}$};
\node[st=violet] (b) at (-2.25,1.0) {$M_{1\cdots N-2,N}$};
\node at (0,1.0) {$\cdots$};
\node[st=violet] (c) at (2.25,1.0) {$M_{124\cdots N}$};
\node[st=violet] (d) at (4.5,1.0) {$M_{13\cdots N}$};
\node[st=purple,very thick] (all) at (0,-0.5) {$M_{12\cdots N}$};
\draw[ed] (12) -- (1); \draw[ed] (13) -- (1); \draw[ed] (14) -- (1); \draw[ed] (1N) -- (1);
\draw[ed] (123) -- (12); \draw[ed] (123) -- (13);
\draw[ed] (124) -- (12); \draw[ed] (124) -- (14);
\draw[ed] (134) -- (13); \draw[ed] (134) -- (14);
\draw[ed] (1MN) -- (1N);
\draw[ed] (v1) -- (123); \draw[ed] (v2) -- (124); \draw[ed] (v3) -- (134); \draw[ed] (v5) -- (1MN);
\draw[ed] (a) -- (v1); \draw[ed] (b) -- (v2); \draw[ed] (c) -- (v4); \draw[ed] (d) -- (v5);
\draw[ed] (all) -- (a); \draw[ed] (all) -- (b); \draw[ed] (all) -- (c); \draw[ed] (all) -- (d);
\draw[edj] (124) -- (1);
\draw[edj] (all) -- (124);
\draw[ed0] (1) -- (0);
\draw[ed0] (13) -- (0);
\draw[ed0] (12) -- (0);
\draw[ed0,rounded corners=6pt] (all.west) -- (-6,-0.5) -- (-6,6.2) -- (0.west);
\end{tikzpicture}
\caption{The extinction network for general $N$, with states ordered by the number of surviving species. Representative transitions are shown: single-species losses (solid), simultaneous losses of several consumers (dashed) and losses of $X_1$, which lead to $M_0$ (dotted). In full, each state connects to every smaller one, giving $3^{N-1}$ transitions}
\label{fig:network}
\end{figure}

The feasible fixed points are the nodes of a directed extinction  network (Fig.~\ref{fig:network}). Although every partial community $M_A$ is a saddle of the full mean-field dynamics, its unstable directions point out of the hyperplane on which the species are extinct. Without immigration the stochastic dynamics never leave that hyperplane, and within it $M_A$ is stable (Lemma~\ref{lem:active}), so it supports a long-lived quasi-stationary distribution. A rare fluctuation can carry the system from $M_A$ to any $M_B$ with $B\subsetneq A$, by the loss of one consumer, of several consumers at once, or of $X_1$, after which the remaining consumers collapse deterministically onto $M_0$. States with $X_1$ extinct and some consumer present are not fixed points. They decay to $M_0$ on a time of order one, which is negligible against the exponentially long residence times, and we include them in the absorbing class $M_0$. Each $M_A$ therefore has $2^{|A|-1}$ outgoing transitions, one for each proper subset of $A$ that contains 1 and one to $M_0$, and the network carries
\begin{equation}\label{eq27}
\sum_{n=1}^{N}\binom{N-1}{n-1}\,2^{\,n-1} = 3^{N-1}
\end{equation}
transitions in total, by the binomial theorem. Their number grows considerably faster than the number of nodes.

Let $P_A(t)$ be the probability that exactly the species in $A$ are present ($1\in A$), and $P_0=1-\sum_AP_A$. Once the system has relaxed into the quasi-stationary distribution about its current state, which takes a time $t_r$ of the order of the inverse slowest relaxation rate of the linearised dynamics, the coarse-grained probabilities obey the rate equations of a continuous-time Markov chain,
\begin{equation}\label{eq28}
\dot P_A = -R_A\,P_A + \sum_{B\supsetneq A} r_{B\to A}\,P_B, \qquad R_A = \sum_{B\subsetneq A} r_{A\to B},
\end{equation}
valid on time scales $t\gg t_r$ provided $t_r\ll1/R_A$ for every $A$. The initial condition $P_{12\cdots N}(0)=1$ describes a community that starts in coexistence. Because \eqref{eq28} is triangular with respect to set inclusion, its eigenvalues are $0$ (the absorbing state) and $-R_A$, which are the slow eigenvalues of the master-equation operator \eqref{eq16}, and its solution is a finite sum of exponentials obtained recursively by integrating factors, as in the Bateman solution of radioactive decay chains \citep{Bateman1910}.

Mean extinction times follow without solving \eqref{eq28} explicitly. Write $F=\{1,\dots,N\}$ for the full community. Define the mean residence times $T_A=\int_0^\infty P_A(t)\,\mathrm dt$. Integrating \eqref{eq28} over $0<t<\infty$, with $P_A(\infty)=0$ for every $A\neq\varnothing$, gives the exact recursion
\begin{equation}\label{eq29}
T_A = \frac{1}{R_A}\Big(\delta_{A,F} + \sum_{B\supsetneq A} r_{B\to A}\,T_B\Big),
\end{equation}
which is solved from the coexistence state downwards. Iterating it gives the explicit solution
\begin{equation}\label{eq:pathsum}
T_A = \frac{\pi_A}{R_A}, \qquad \pi_A = \sum_{F=A_0\supsetneq A_1\supsetneq\dots\supsetneq A_m=A}\ \prod_{i=0}^{m-1}\frac{r_{A_i\to A_{i+1}}}{R_{A_i}}, \qquad \pi_F = 1,
\end{equation}
where the sum runs over every chain of states from the full community down to $A$. Each factor $r_{A_i\to A_{i+1}}/R_{A_i}$ is the probability that the community leaves $A_i$ for $A_{i+1}$, so $\pi_A$ is the probability that the community ever occupies $M_A$, and $1/R_A$ is its mean residence time there. The same products give the probability of any complete extinction route $F\to A_1\to\dots\to M_0$.

The survival probability of species $k$ is $\sum P_A$ over the active sets that contain $k$, and a mean first-passage time is the time integral of the corresponding survival probability. The mean time to extinction of $X_k$, the mean time until every species in a set $K$ has been lost, and the mean time to total extinction are therefore
\begin{equation}\label{eq30}
\tau_k = \sum_{A\ni k}\frac{\pi_A}{R_A}, \qquad \tau_K = \sum_{A\cap K\neq\varnothing}\frac{\pi_A}{R_A}, \qquad \tau_{\mathrm{tot}} = \tau_1 = \sum_{A\neq\varnothing}\frac{\pi_A}{R_A},
\end{equation}
with all sums restricted to admissible active sets ($1\in A$). These are the $N$-species generalisations of the four three-species times of Section~\ref{sec:three-rates}, $\tau_k$ generalises $\tau_2$ and $\tau_3$, $\tau_K$ with $K=\{2,\dots,N\}$ is the mean time until all consumers have gone, which generalises $\tau_{23}$, and $\tau_{\mathrm{tot}}$ coincides with the extinction time of the basal species, because the consumers cannot outlive it. The mean time to the first extinction of any species is $T_F=1/R_F$. Equations \eqref{eq:pathsum} and \eqref{eq30} express every mean extinction time as a rational function of the $3^{N-1}$ transition rates.

\subsection{Combinatorial scaling}

Table~\ref{tab:scaling} summarises how the numbers of fixed points, transitions and existence conditions develop with $N$. The extinction network grows as $3^{N-1}$, whereas the reduced model has only $(N^2+N+4)/2$ parameters, so for $N=50$ some $10^{23}$ transitions are controlled by $1\,277$ parameters. A small set of parameters therefore fixes the behaviour of a large food web, but identifying which transitions and conditions are essential becomes the central difficulty as $N$ grows.

\begin{table}[ht]
\centering
\caption{Combinatorial structure of the $N$-species model. Entries for $N=50$ and $N=100$ are rounded to two significant figures}
\label{tab:scaling}
\setlength{\tabcolsep}{3.5pt}
\begin{tabular}{@{}>{\raggedright\arraybackslash}p{3.0cm}cccccc@{}}
\toprule
Quantity & General $N$ & $N=3$ & $N=5$ & $N=10$ & $N=50$ & $N=100$ \\
\midrule
Feasible fixed points & $2^{N-1}+1$ & 5 & 17 & 513 & $5.6\times10^{14}$ & $6.3\times10^{29}$ \\
Transitions & $3^{N-1}$ & 9 & 81 & 19\,683 & $2.4\times10^{23}$ & $1.7\times10^{47}$ \\
Fluctuational fixed points & $3^{N-1}-2^{N-1}+1$ & 6 & 66 & 19\,172 & $2.4\times10^{23}$ & $1.7\times10^{47}$ \\
Parameters, full model \eqref{eq1} & $N(N+2)$ & 15 & 35 & 120 & 2\,600 & 10\,200 \\
Parameters, reduced model \eqref{eq2}--\eqref{eq3} & $(N^2+N+4)/2$ & 8 & 17 & 57 & 1\,277 & 5\,052 \\
Coordinate positivity conditions & $(N+1)\,2^{N-2}$ & 8 & 48 & 2\,816 & $1.4\times10^{16}$ & $3.2\times10^{31}$ \\
Non-trivial positivity conditions & $(N+1)\,2^{N-2}-(N-1)$ & 6 & 44 & 2\,807 & $1.4\times10^{16}$ & $3.2\times10^{31}$ \\
\bottomrule
\end{tabular}
\end{table}

\section{The three-species omnivory model}\label{sec:three}

We now specialise the general theory to $N=3$, the first case with omnivory, where every fixed point, eigenvalue and existence condition of Section~\ref{sec:general} can be written in closed form.

\subsection{Model}

\begin{figure}[ht]
\centering
\resizebox{\textwidth}{!}{%
\begin{tikzpicture}[
  sp/.style={circle,draw=#1!70!black,fill=#1!15,thick,minimum size=8mm,inner sep=0pt,font=\small},
  ed/.style={-latex,thick,blue!70!black},
  edo/.style={-latex,thick,dashed,orange!90!black},
  edc/.style={-latex,thick,densely dotted,red!75!black},
  lab/.style={font=\small,align=center,anchor=north}]
\begin{scope}
\node[sp=green] (1) at (0,0) {$X_1$};
\node[sp=teal] (2) at (-1.0,1.4) {$X_2$};
\node[sp=cyan] (3) at (0,2.8) {$X_3$};
\draw[ed] (1) -- (2); \draw[ed] (2) -- (3); \draw[edo] (1) -- (3);
\node[lab] at (-0.35,-0.65) {(a)};
\end{scope}
\begin{scope}[xshift=3.0cm]
\node[sp=green] (1) at (0,0) {$X_1$};
\node[sp=teal] (2) at (-1.0,1.4) {$X_2$};
\node[sp=cyan] (3) at (0,2.8) {$X_3$};
\draw[ed] (1) -- (2); \draw[ed] (2) -- (3);
\node[lab] at (-0.35,-0.65) {(b)};
\end{scope}
\begin{scope}[xshift=6.0cm]
\node[sp=green] (1) at (0,0) {$X_1$};
\node[sp=teal] (2) at (-1.0,1.4) {$X_2$};
\node[sp=cyan] (3) at (0,2.8) {$X_3$};
\draw[ed] (2) -- (3); \draw[ed] (1) -- (3);
\node[lab] at (-0.35,-0.65) {(c)};
\end{scope}
\begin{scope}[xshift=9.0cm]
\node[sp=green] (1) at (0,0) {$X_1$};
\node[sp=teal] (2) at (-1.0,1.4) {$X_2$};
\node[sp=cyan] (3) at (0,2.8) {$X_3$};
\draw[ed] (1) -- (2); \draw[ed] (1) -- (3);
\node[lab] at (-0.35,-0.65) {(d)};
\end{scope}
\begin{scope}[xshift=12.0cm]
\node[sp=green] (1) at (0,0) {$X_1$};
\node[sp=teal] (2) at (-1.0,1.4) {$X_2$};
\node[sp=cyan] (3) at (0,2.8) {$X_3$};
\draw[ed] (1) -- (2); \draw[ed] (2) -- (3); \draw[edc] (3) -- (1);
\node[lab] at (-0.35,-0.65) {(e)};
\end{scope}
\end{tikzpicture}%
}
\caption{The five connected three-species food-web topologies, A086345(3)~$=5$, with arrows pointing from prey to predator. (a)~The omnivory module studied in this work, including the omnivorous link (dashed). (b)~The food chain, $a_{13}=0$. (c)~A shared predator, $a_{12}=0$, in which $X_3$ preys on both $X_1$ and $X_2$. (d)~A shared prey, $a_{23}=0$, in which $X_2$ and $X_3$ compete for $X_1$. (e)~The cyclic web, in which $X_3$ preys on $X_1$ (dotted)}
\label{fig:omnivory}
\end{figure}
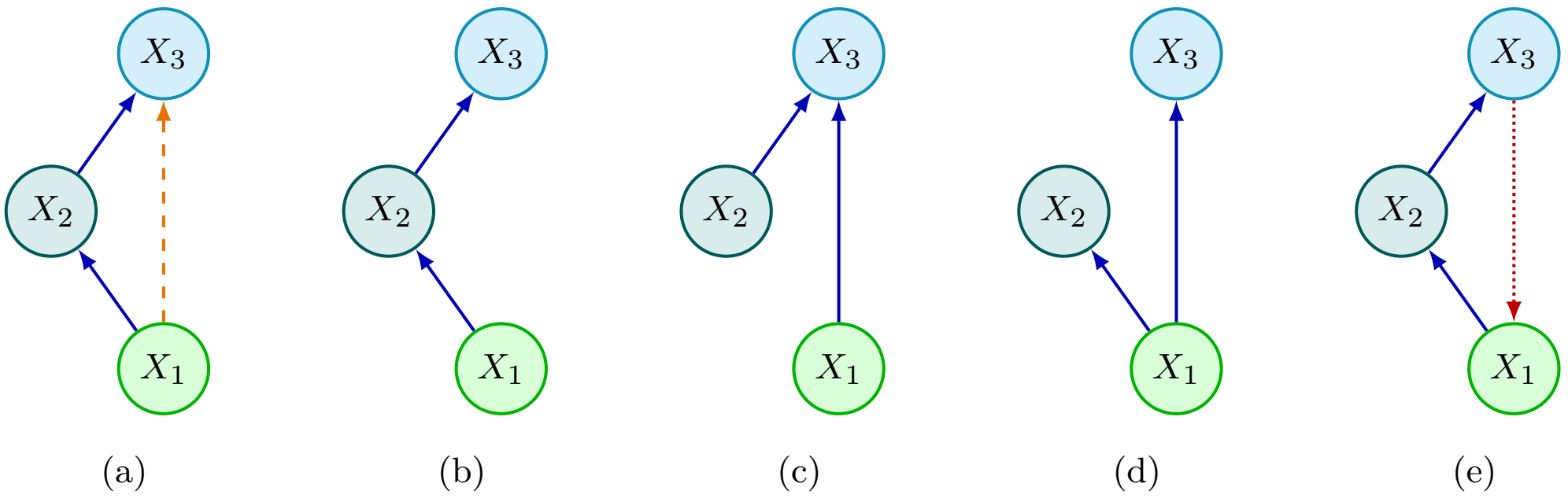

The basal species $X_1$ is consumed by the intermediate species $X_2$, and the apex species $X_3$ preys both on $X_2$ and directly on $X_1$ (Fig.~\ref{fig:omnivory}). This is the Lotka--Volterra intraguild-predation module \citep{HoltPolis1997}, in which $X_2$ and $X_3$ compete for $X_1$ while $X_3$ also consumes $X_2$. For $N=3$ the full mean-field equations \eqref{eq1} read
\begin{equation}\label{eq31}
\begin{aligned} \dot x_1 &= B_1x_1 - a_{11}x_1^2 - a_{12}x_1x_2 - a_{13}x_1x_3, \\ \dot x_2 &= B_2x_2 - a_{22}x_2^2 + e_{12}a_{12}x_1x_2 - a_{23}x_2x_3, \\ \dot x_3 &= B_3x_3 - a_{33}x_3^2 + e_{13}a_{13}x_1x_3 + e_{23}a_{23}x_2x_3, \end{aligned} \qquad B_k \equiv b_k - d_k,
\end{equation}
with fifteen parameters. Generically, \eqref{eq31} has $2^3=8$ fixed points. Four lie at the origin or on the axes,
\begin{equation}\label{eq32}
(0,0,0), \qquad \Big(\frac{B_1}{a_{11}},0,0\Big), \qquad \Big(0,\frac{B_2}{a_{22}},0\Big), \qquad \Big(0,0,\frac{B_3}{a_{33}}\Big),
\end{equation}
three lie in the coordinate planes and describe the coexistence of $X_2$ with $X_3$, of $X_1$ with $X_3$, and of $X_1$ with $X_2$,
\begin{equation}\label{eq33}
\begin{gathered}
\Big(0,\ \frac{B_2a_{33}-B_3a_{23}}{a_{22}a_{33}+e_{23}a_{23}^2},\ \frac{a_{22}B_3+e_{23}a_{23}B_2}{a_{22}a_{33}+e_{23}a_{23}^2}\Big), \\
\Big(\frac{B_1a_{33}-B_3a_{13}}{a_{11}a_{33}+e_{13}a_{13}^2},\ 0,\ \frac{a_{11}B_3+e_{13}a_{13}B_1}{a_{11}a_{33}+e_{13}a_{13}^2}\Big), \\
\Big(\frac{B_1a_{22}-B_2a_{12}}{a_{11}a_{22}+e_{12}a_{12}^2},\ \frac{a_{11}B_2+e_{12}a_{12}B_1}{a_{11}a_{22}+e_{12}a_{12}^2},\ 0\Big),
\end{gathered}
\end{equation}
and the interior fixed point solves
\begin{equation}\label{eq34}
\begin{pmatrix} a_{11} & a_{12} & a_{13} \\ -e_{12}a_{12} & a_{22} & a_{23} \\ -e_{13}a_{13} & -e_{23}a_{23} & a_{33} \end{pmatrix}\begin{pmatrix} x_1 \\ x_2 \\ x_3 \end{pmatrix} = \begin{pmatrix} B_1 \\ B_2 \\ B_3 \end{pmatrix}.
\end{equation}
Under the reductions of Section~\ref{sec:reduced} ($b_2=b_3=0$, $a_{22}=a_{33}=0$ and $e_{ij}=1$), the $X_2$ and $X_3$ axial points of \eqref{eq32} cease to exist, because without self-limitation and with $B_2=-d_2<0$ and $B_3=-d_3<0$ the consumer equations have no positive axial root. The $X_2$--$X_3$ fixed point of \eqref{eq33} becomes $(0,\,d_3/a_{23},\,-d_2/a_{23})$, whose third coordinate is negative, so this point is never feasible. The determinant of \eqref{eq34} correspondingly reduces to $a_{11}a_{23}^2$. Five fixed points remain, in agreement with \eqref{eq5}. The reduced equations are
\begin{equation}\label{eq35}
\begin{aligned} \dot x_1 &= x_1\big(B_1 - a_{11}x_1 - a_{12}x_2 - a_{13}x_3\big), \\ \dot x_2 &= x_2\big(-d_2 + a_{12}x_1 - a_{23}x_3\big), \\ \dot x_3 &= x_3\big(-d_3 + a_{13}x_1 + a_{23}x_2\big), \end{aligned}
\end{equation}
with eight parameters: $b_1$, $d_1$, $d_2$, $d_3$, $a_{11}$, $a_{12}$, $a_{13}$ and $a_{23}$. A strict hierarchy on three species admits four connected topologies, A001187(3)~$=4$ (Appendix~\ref{app:graphs}): the full omnivory module and the three obtained by deleting a single link, which are analysed in Section~\ref{sec:reducedtop}. Without a hierarchy a fifth, cyclic topology $X_1\to X_2\to X_3\to X_1$ is also possible, completing the A086345(3)~$=5$ connected oriented topologies. Its WKB extinction problem has been treated separately for the May--Leonard model of three cyclically competing species \citep{Serrao2021}.

\subsection{Fixed points and existence conditions}\label{sec:three-fp}

It is convenient to introduce
\begin{equation}\label{eq36}
\begin{gathered}
u = B_1a_{12} - a_{11}d_2, \qquad v = B_1a_{13} - a_{11}d_3, \qquad w = a_{13}d_2 - a_{12}d_3, \\
a_{13}\,u - a_{12}\,v = -a_{11}\,w.
\end{gathered}
\end{equation}
The five fixed points of \eqref{eq35} are $M_0=(0,0,0)$, $M_1=(B_1/a_{11},0,0)$ and
\begin{equation}\label{eq37}
M_{12} = \Big(\frac{d_2}{a_{12}},\ \frac{u}{a_{12}^2},\ 0\Big), \qquad M_{13} = \Big(\frac{d_3}{a_{13}},\ 0,\ \frac{v}{a_{13}^2}\Big),
\end{equation}
\begin{equation}\label{eq38}
M_{123}:\quad x_1^{*} = \frac{a_{23}B_1 + w}{a_{11}a_{23}}, \qquad x_2^{*} = -\,\frac{a_{23}v + a_{13}w}{a_{11}a_{23}^2}, \qquad x_3^{*} = \frac{a_{23}u + a_{12}w}{a_{11}a_{23}^2},
\end{equation}
which is \eqref{eq10} with $j=2$ and $k=3$. Eliminating $w$ with the identity in \eqref{eq36} gives
\begin{equation}\label{eq39}
a_{11}^2a_{23}^2\,x_2^{*} = a_{13}^2\,u - \big(a_{11}a_{23} + a_{12}a_{13}\big)\,v, \qquad a_{11}^2a_{23}^2\,x_3^{*} = \big(a_{11}a_{23} - a_{12}a_{13}\big)\,u + a_{12}^2\,v.
\end{equation}
By Lemma~\ref{lem:establishment}, $x_3^*$ has the sign of the passive eigenvalue of $X_3$ at $M_{12}$, and $x_2^*$ that of $X_2$ at $M_{13}$.

\begin{proposition}\label{prop:existence}
All five fixed points are feasible if and only if
\begin{equation}\label{eq40}
a_{13}B_1 > a_{11}d_3,
\end{equation}
\begin{equation}\label{eq41}
d_3\big(a_{11}a_{23} + a_{12}a_{13}\big) > a_{13}a_{23}B_1 + a_{13}^2d_2,
\end{equation}
\begin{equation}\label{eq42}
a_{12}a_{23}B_1 + a_{12}a_{13}d_2 > a_{11}a_{23}d_2 + a_{12}^2d_3.
\end{equation}
\end{proposition}

\begin{proof}
Conditions \eqref{eq40}, \eqref{eq41} and \eqref{eq42} are $v>0$, $x_2^*>0$ and $x_3^*>0$, so they are necessary. Conversely, \eqref{eq40} implies $B_1>0$, so $M_1$ is feasible. If $u\le0$, the first relation in \eqref{eq39} with $x_2^*>0$ would give $(a_{11}a_{23}+a_{12}a_{13})v<a_{13}^2u\le0$, contradicting \eqref{eq40}, hence $u>0$ and $M_{12}$ is feasible. Finally, $x_3^*>0$ means $a_{12}w>-a_{23}u$, so $a_{12}(a_{23}B_1+w)>a_{23}(a_{12}B_1-u)=a_{11}a_{23}d_2>0$, and $x_1^*>0$.
\end{proof}

The six non-trivial positivity conditions of Section~\ref{sec:fixedpoints} therefore reduce to three. Conditions \eqref{eq41} and \eqref{eq42} 	express mutual establishment. Each consumer can establish itself in the community formed by $X_1$ and the other consumer, the familiar form of coexistence conditions in intraguild-predation theory \citep{HoltPolis1997}.

\subsection{Stability}\label{sec:three-stab}

We now make each case of Theorem~\ref{thm:saddle} explicit, using the Jacobian \eqref{eq11} and the criteria of Appendix~\ref{app:stability}.

\paragraph{Total extinction $M_0$.} The Jacobian is $\operatorname{diag}(B_1,-d_2,-d_3)$, so $M_0$ is a saddle with one unstable direction.

\paragraph{Basal state $M_1$.} The Jacobian is upper-triangular with eigenvalues $-B_1<0$, $f_2=u/a_{11}>0$ and $f_3=v/a_{11}>0$, so $M_1$ is a saddle with two unstable directions, one for each consumer that can establish.

\paragraph{Two-species states $M_{12}$ and $M_{13}$.} Writing $(x_1,x_2,0)$ and $(x_1,0,x_3)$ for the coordinates \eqref{eq37}, the Jacobians are
\begin{equation}\label{eq43}
\mathbf J_{12} = \begin{pmatrix} -a_{11}x_1 & -d_2 & -a_{13}x_1 \\ a_{12}x_2 & 0 & -a_{23}x_2 \\ 0 & 0 & f_3 \end{pmatrix}, \qquad \mathbf J_{13} = \begin{pmatrix} -a_{11}x_1 & -a_{12}x_1 & -d_3 \\ 0 & f_2 & 0 \\ a_{13}x_3 & a_{23}x_3 & 0 \end{pmatrix},
\end{equation}
with passive eigenvalues
\begin{equation}\label{eq44}
f_3^{(12)} = \frac{a_{11}a_{23}^2}{a_{12}^2}\,x_3^{*} > 0, \qquad f_2^{(13)} = \frac{a_{11}a_{23}^2}{a_{13}^2}\,x_2^{*} > 0,
\end{equation}
in agreement with Lemma~\ref{lem:establishment}, since $\det \mathbf L_{\{12\}}=-a_{12}^2$, $\det \mathbf L_{\{13\}}=-a_{13}^2$ and $\det \mathbf L_{\{123\}}=a_{11}a_{23}^2$. The active blocks have characteristic equations
\begin{equation}\label{eq45}
a_{12}\lambda^2 + a_{11}d_2\,\lambda + d_2\,u = 0, \qquad a_{13}\lambda^2 + a_{11}d_3\,\lambda + d_3\,v = 0,
\end{equation}
whose roots have negative sum and positive product, so both lie in the left half-plane. Each two-species state is therefore a saddle with a single unstable direction, the establishment of the missing consumer.

\paragraph{Coexistence $M_{123}$.} The Jacobian
\begin{equation}\label{eq46}
\mathbf J_{123} = \begin{pmatrix} -a_{11}x_1^{*} & -a_{12}x_1^{*} & -a_{13}x_1^{*} \\ a_{12}x_2^{*} & 0 & -a_{23}x_2^{*} \\ a_{13}x_3^{*} & a_{23}x_3^{*} & 0 \end{pmatrix}
\end{equation}
has characteristic polynomial $\lambda^3+c_2\lambda^2+c_1\lambda+c_0$ with
\begin{equation}\label{eq47}
\begin{aligned}
c_2 &= a_{11}x_1^{*}, & c_1 &= a_{12}^2x_1^{*}x_2^{*} + a_{13}^2x_1^{*}x_3^{*} + a_{23}^2x_2^{*}x_3^{*}, \\
c_0 &= a_{11}a_{23}^2\,x_1^{*}x_2^{*}x_3^{*}, & c_1c_2 - c_0 &= a_{11}x_1^{*2}\big(a_{12}^2x_2^{*} + a_{13}^2x_3^{*}\big).
\end{aligned}
\end{equation}
All three Routh-Hurwitz quantities $c_2$, $c_0$ and $c_1c_2-c_0$ are positive, so every eigenvalue has negative real part and $M_{123}$ is asymptotically stable. Its relaxation time $t_r=1/\min|\operatorname{Re}\lambda|$ sets the time needed to settle into the quasi-stationary distribution in Section~\ref{sec:three-rates}. The stronger, global statement also holds.

\begin{proposition}[Global stability]\label{prop:global}
Under \eqref{eq40}--\eqref{eq42}, every trajectory of \eqref{eq35} starting in the open positive orthant converges to $M_{123}$.
\end{proposition}

\begin{proof}
By \eqref{eq15}, $\dot V=-a_{11}(x_1-x_1^*)^2\le0$, and $V$ is radially unbounded on the open orthant, so by LaSalle's invariance principle every trajectory approaches the largest invariant set contained in $\{x_1=x_1^*\}$. On such a set $\dot x_1=0$, which gives $a_{12}(x_2-x_2^*)+a_{13}(x_3-x_3^*)=0$, while $\dot x_2=-a_{23}x_2(x_3-x_3^*)$ and $\dot x_3=a_{23}x_3(x_2-x_2^*)$. Differentiating the constraint and eliminating $x_2-x_2^*$ yields $(a_{23}/a_{12})(x_3-x_3^*)(a_{12}^2x_2+a_{13}^2x_3)=0$. The bracket is positive, so $x_3=x_3^*$, and then $x_2=x_2^*$. The invariant set is the single point $M_{123}$.
\end{proof}

Under the assumptions of Section~\ref{sec:reduced} and the existence conditions \eqref{eq40}--\eqref{eq42}, the omnivory module therefore settles into coexistence from every initial state in which all three species are present, and every species loss must be fluctuation-driven. When these conditions fail, the attractor lies on a boundary of the positive orthant (Section~\ref{sec:reducedtop}).

\subsection{Hamiltonian and fluctuational fixed points}

For $N=3$ the Hamiltonian \eqref{eq17} is
\begin{equation}\label{eq48}
\begin{aligned} H ={}& b_1x_1\big(\e^{p_1}-1\big) + \big(d_1x_1 + a_{11}x_1^2\big)\big(\e^{-p_1}-1\big) + d_2x_2\big(\e^{-p_2}-1\big) \\ &+ d_3x_3\big(\e^{-p_3}-1\big) + a_{12}x_1x_2\big(\e^{p_2-p_1}-1\big) \\ &+ a_{13}x_1x_3\big(\e^{p_3-p_1}-1\big) + a_{23}x_2x_3\big(\e^{p_3-p_2}-1\big), \end{aligned}
\end{equation}
and Hamilton's equations are
\begin{equation}\label{eq49}
\begin{aligned} \dot x_1 &= b_1x_1\e^{p_1} - \big(d_1x_1 + a_{11}x_1^2\big)\e^{-p_1} - a_{12}x_1x_2\,\e^{p_2-p_1} - a_{13}x_1x_3\,\e^{p_3-p_1}, \\ \dot x_2 &= -d_2x_2\e^{-p_2} + a_{12}x_1x_2\,\e^{p_2-p_1} - a_{23}x_2x_3\,\e^{p_3-p_2}, \\ \dot x_3 &= -d_3x_3\e^{-p_3} + a_{13}x_1x_3\,\e^{p_3-p_1} + a_{23}x_2x_3\,\e^{p_3-p_2}, \\ \dot p_1 &= -\big[b_1\big(\e^{p_1}-1\big) + \big(d_1 + 2a_{11}x_1\big)\big(\e^{-p_1}-1\big) \\ &\qquad + a_{12}x_2\big(\e^{p_2-p_1}-1\big) + a_{13}x_3\big(\e^{p_3-p_1}-1\big)\big], \\ \dot p_2 &= -\big[d_2\big(\e^{-p_2}-1\big) + a_{12}x_1\big(\e^{p_2-p_1}-1\big) + a_{23}x_3\big(\e^{p_3-p_2}-1\big)\big], \\ \dot p_3 &= -\big[d_3\big(\e^{-p_3}-1\big) + a_{13}x_1\big(\e^{p_3-p_1}-1\big) + a_{23}x_2\big(\e^{p_3-p_2}-1\big)\big]. \end{aligned}
\end{equation}
Applying \eqref{eq24} and \eqref{eq25} gives all six fluctuational fixed points, written as $(x_1,x_2,x_3;\allowbreak p_1,p_2,p_3)$.
\begin{equation}\label{eq50}
F_0 = \Big(0,0,0;\ \ln\frac{d_1}{b_1},\,0,\,0\Big).
\end{equation}
At $M_1$ both consumers are extinct, and their momenta take the values
\begin{equation}\label{eq51}
p_2^{*} = \ln\frac{a_{11}d_2}{a_{12}B_1}, \qquad p_3^{*} = \ln\frac{a_{11}d_3}{a_{13}B_1},
\end{equation}
independently, which yields $F_1^{(2)}=(B_1/a_{11},0,0;0,p_2^*,0)$, $F_1^{(3)}=(B_1/a_{11},0,0;0,0,p_3^*)$ and $F_1^{(23)}=(B_1/a_{11},0,0;0,p_2^*,p_3^*)$. Both logarithms in \eqref{eq51} are negative, because $u>0$ and $v>0$. At the two-species states only one species is extinct:
\begin{equation}\label{eq52}
F_{12}^{(3)}:\ \ p_3^{*} = \ln\frac{a_{12}^2d_3}{a_{12}a_{23}B_1 + a_{12}a_{13}d_2 - a_{11}a_{23}d_2} = \ln\frac{d_3}{d_3 + f_3^{(12)}},
\end{equation}
\begin{equation}\label{eq53}
F_{13}^{(2)}:\ \ p_2^{*} = \ln\frac{a_{13}^2d_2 + a_{13}a_{23}B_1 - a_{11}a_{23}d_3}{a_{12}a_{13}d_3} = \ln\Big(1 - \frac{f_2^{(13)}}{a_{12}x_1}\Big),
\end{equation}
where the positions are those of $M_{12}$ and $M_{13}$, $x_1=d_3/a_{13}$ in \eqref{eq53}, and all other momenta vanish. Both momenta are negative by \eqref{eq44}. The coexistence state has no extinct species and hence no fluctuational fixed point. Every extinction route out of $M_{123}$ ends on one of the invariant hyperplanes.

Table~\ref{tab:transitions} lists the nine transitions of the extinction network together with the fluctuational point at which each instanton terminates. Following Section~\ref{sec:fluct}, the end point must carry non-zero momenta for the species lost in the transition. For $M_{12}\to M_1$ and $M_{13}\to M_1$ there are two candidates, which differ in whether the momentum of the consumer that was already extinct also ends on its fluctuational root. This must be settled by comparing actions in the full six-dimensional phase space.

\begin{table}[ht]
\centering
\caption{Transitions of the three-species extinction network}
\label{tab:transitions}
\begin{tabular}{llll}
\toprule
Transition & Rate & Species lost & Instanton end point \\
\midrule
$M_1\to M_0$ & $r_{1\to0}$ & $X_1$ & $F_0$ \\
$M_{12}\to M_1$ & $r_{12\to1}$ & $X_2$ & $F_1^{(2)}$ or $F_1^{(23)}$ \\
$M_{12}\to M_0$ & $r_{12\to0}$ & $X_1$, then $X_2$ & $F_0$ \\
$M_{13}\to M_1$ & $r_{13\to1}$ & $X_3$ & $F_1^{(3)}$ or $F_1^{(23)}$ \\
$M_{13}\to M_0$ & $r_{13\to0}$ & $X_1$, then $X_3$ & $F_0$ \\
$M_{123}\to M_{12}$ & $r_{123\to12}$ & $X_3$ & $F_{12}^{(3)}$ \\
$M_{123}\to M_{13}$ & $r_{123\to13}$ & $X_2$ & $F_{13}^{(2)}$ \\
$M_{123}\to M_1$ & $r_{123\to1}$ & $X_2$ and $X_3$ together & $F_1^{(23)}$ \\
$M_{123}\to M_0$ & $r_{123\to0}$ & $X_1$, then $X_2$ and $X_3$ & $F_0$ \\
\bottomrule
\end{tabular}
\end{table}

\subsection{Occupation probabilities and mean extinction times}\label{sec:three-rates}

For $N=3$ the rate equations \eqref{eq28} read
\begin{equation}\label{eq54}
\begin{aligned} \dot P_{123} &= -R_{123}\,P_{123}, \\ \dot P_{12} &= -R_{12}\,P_{12} + r_{123\to12}\,P_{123}, \\ \dot P_{13} &= -R_{13}\,P_{13} + r_{123\to13}\,P_{123}, \\ \dot P_{1} &= -r_{1\to0}\,P_{1} + r_{12\to1}\,P_{12} + r_{13\to1}\,P_{13} + r_{123\to1}\,P_{123}, \end{aligned}
\end{equation}
with $P_0=1-P_1-P_{12}-P_{13}-P_{123}$ and total outflow rates
\begin{equation}\label{eq55}
\begin{gathered}
R_{123} = r_{123\to12} + r_{123\to13} + r_{123\to1} + r_{123\to0}, \\
R_{12} = r_{12\to1} + r_{12\to0}, \qquad R_{13} = r_{13\to1} + r_{13\to0}.
\end{gathered}
\end{equation}
For a community that starts in coexistence, $P_{123}(0)=1$, and for distinct rates, integration by integrating factors gives
\begin{equation}\label{eq56}
\begin{gathered}
P_{123} = \e^{-R_{123}t}, \qquad P_{12} = \frac{r_{123\to12}}{R_{12}-R_{123}}\big(\e^{-R_{123}t} - \e^{-R_{12}t}\big), \\
P_{13} = \frac{r_{123\to13}}{R_{13}-R_{123}}\big(\e^{-R_{123}t} - \e^{-R_{13}t}\big),
\end{gathered}
\end{equation}
\begin{equation}\label{eq57}
\begin{aligned}
P_1 ={}& \frac{C_{123}}{r_{1\to0}-R_{123}}\big(\e^{-R_{123}t} - \e^{-r_{1\to0}t}\big) - \frac{C_{12}}{r_{1\to0}-R_{12}}\big(\e^{-R_{12}t} - \e^{-r_{1\to0}t}\big) \\
&- \frac{C_{13}}{r_{1\to0}-R_{13}}\big(\e^{-R_{13}t} - \e^{-r_{1\to0}t}\big),
\end{aligned}
\end{equation}
with
\begin{equation}\label{eq58}
C_{12} = \frac{r_{123\to12}\,r_{12\to1}}{R_{12}-R_{123}}, \qquad C_{13} = \frac{r_{123\to13}\,r_{13\to1}}{R_{13}-R_{123}}, \qquad C_{123} = r_{123\to1} + C_{12} + C_{13}.
\end{equation}
The recursion \eqref{eq29} gives the mean residence times
\begin{equation}\label{eq59}
\begin{gathered}
T_{123} = \frac{1}{R_{123}}, \qquad T_{12} = \frac{r_{123\to12}}{R_{12}R_{123}}, \qquad T_{13} = \frac{r_{123\to13}}{R_{13}R_{123}}, \\
T_1 = \frac{1}{r_{1\to0}R_{123}}\Big(r_{123\to1} + \frac{r_{123\to12}\,r_{12\to1}}{R_{12}} + \frac{r_{123\to13}\,r_{13\to1}}{R_{13}}\Big),
\end{gathered}
\end{equation}
and \eqref{eq30}, with $k=2$, $k=3$, $K=\{2,3\}$ and $A\neq\varnothing$, gives the mean extinction times of $X_2$, of $X_3$, of both consumers, and of the whole community:
\begin{equation}\label{eq60}
\begin{gathered}
\tau_2 = \frac{1}{R_{123}}\Big(1 + \frac{r_{123\to12}}{R_{12}}\Big), \qquad \tau_3 = \frac{1}{R_{123}}\Big(1 + \frac{r_{123\to13}}{R_{13}}\Big), \\
\tau_{23} = \frac{1}{R_{123}}\Big(1 + \frac{r_{123\to12}}{R_{12}} + \frac{r_{123\to13}}{R_{13}}\Big), \qquad \tau_{\mathrm{tot}} = \tau_{23} + T_1.
\end{gathered}
\end{equation}
These four time scales, expressed through the nine transition rates, are the principal quantitative predictions of the theory for the omnivory module. They require the separation of time scales $t_r\ll\min(1/R_{123},1/R_{12},1/R_{13},1/r_{1\to0})$.

Every realisation ends in $M_0$, but the sequence of states it visits on the way is random. A history that starts in coexistence can be in one of twelve distinct situations: no loss (1); $M_{12}$ or $M_{13}$ reached directly (1 each); $M_1$ reached directly or through either two-species state (3); and $M_0$ reached along one of six routes, namely $123\to0$, $123\to1\to0$, $123\to12\to0$, $123\to12\to1\to0$, $123\to13\to0$ and $123\to13\to1\to0$. The probability of a complete route $A_0\to A_1\to\dots\to M_0$ is the product of branching ratios $r_{A_i\to A_{i+1}}/R_{A_i}$, which generalises the comparison of extinction routes in \citet{GottesmanMeerson2012}.

As a consistency check on the master equation \eqref{eq16}, summing it over the appropriate states gives exact identities in which the competition terms cancel, for example
\begin{equation}\label{eq61}
\begin{gathered}
\frac{\mathrm d}{\mathrm dt}\sum_{X_1\ge0} P(X_1,0,0) = \sum_{X_1\ge0}\big[d_2\,P(X_1,1,0) + d_3\,P(X_1,0,1)\big], \\
\frac{\mathrm d}{\mathrm dt}P(0,0,0) = d_1P(1,0,0) + d_2P(0,1,0) + d_3P(0,0,1).
\end{gathered}
\end{equation}

\subsection{Single-species benchmark}

The transition $M_1\to M_0$ is one-dimensional and can be solved exactly. On the $X_1$ axis the zero-energy condition $H=0$ has the fluctuational branch $\e^{p_1}=(d_1+a_{11}x_1)/b_1$, along which the action is
\begin{equation}\label{eq62}
S_{1\to0} = \int_0^{B_1/a_{11}} \ln\frac{b_1}{d_1 + a_{11}x_1}\;\mathrm dx_1 = \frac{1}{a_{11}}\Big(B_1 + d_1\ln\frac{d_1}{b_1}\Big) > 0.
\end{equation}
The pre-exponential factor follows from the exact mean first-passage time of the one-step process with birth rate $b_1X_1$ and death rate $d_1X_1+a_{11}X_1(X_1-1)/\Omega$, evaluated with the Euler-Maclaurin formula and Laplace's method:
\begin{equation}\label{eq63}
r_{1\to0} \simeq \frac{B_1^2}{b_1}\sqrt{\frac{\Omega\,d_1}{2\pi a_{11}}}\;\exp\big(-\Omega\,S_{1\to0}\big).
\end{equation}
 Equation \eqref{eq63} is the benchmark against which the genuinely multi-species rates are compared.

\subsection{Optimal extinction paths}\label{sec:numerics}

Every other rate in Table~\ref{tab:transitions} requires an instanton in the six-dimensional phase space of \eqref{eq49}. At $M_A$ the unstable manifold is three-dimensional (Section~\ref{sec:wkb}), as is the stable manifold of the target $F_B^{(S)}$. Both lie in the five-dimensional zero-energy surface, so their intersection is generically one-dimensional, and the instanton is isolated up to a shift in time. We compute it as a two-point boundary-value problem, imposing at each end that the deviation lie in the appropriate eigenspace of the linearised flow, select the minimum-action solution among candidates obtained from many initial seeds, and require the trajectory and the action \eqref{eq22} to be unchanged under mesh refinement.

\paragraph{Parameters and fixed points.} We use $b_1=1.0$, $d_1=0.8$, $a_{11}=0.05$, $d_2=0.18$, $d_3=0.36$, $a_{12}=0.15$, $a_{13}=0.15$ and $a_{23}=0.3$, which satisfy \eqref{eq40}--\eqref{eq42}. Then $B_1=0.2$, $u=0.021$, $v=0.012$ and $w=-0.027$, and the five fixed points are
\begin{equation}\label{eq:numfp}
\begin{gathered}
M_1 = (4,\,0,\,0), \qquad M_{12} = (1.2,\,0.933,\,0), \\
M_{13} = (2.4,\,0,\,0.533), \qquad M_{123} = (2.2,\,0.1,\,0.5).
\end{gathered}
\end{equation}
The eigenvalues of the active blocks are $-0.0474\pm0.1747\,i$ and $-0.0151$ at $M_{123}$, $-0.0300\pm0.1559\,i$ at $M_{12}$, and $-0.0600\pm0.1587\,i$ at $M_{13}$, so all three are stable foci within their own hyperplanes; the slowest rate gives $t_r\approx66$. The passive eigenvalues \eqref{eq44} are $f_3^{(12)}=0.1$ and $f_2^{(13)}=0.02$. The fluctuational momenta \eqref{eq50}--\eqref{eq53} are $p_1=\ln0.8=-0.223$ at $F_0$; $p_2^*=\ln0.3=-1.204$ and $p_3^*=\ln0.6=-0.511$ at $F_1$; $p_3^*=\ln(0.36/0.46)=-0.245$ at $F_{12}^{(3)}$; and $p_2^*=-0.057$ at $F_{13}^{(2)}$. The end points of the computed instantons agree with these closed forms.

\paragraph{Actions.} Table~\ref{tab:actions} lists the action of every transition. The action of $M_1\to M_0$ is the exact result \eqref{eq62}, the other eight are numerical.

\begin{table}[ht]
\centering
\caption{Instanton actions for the nine transitions of the three-species extinction network, for the parameters of Section~\ref{sec:numerics}. To leading order the rate of each transition is $r\asymp\exp(-\Omega S)$}
\label{tab:actions}
\begin{tabular}{llll}
\toprule
Transition & Species lost & Instanton end point & Action $S$ \\
\midrule
$M_1\to M_0$ & $X_1$ & $F_0$ & 0.429703  \\
$M_{12}\to M_1$ & $X_2$ & $F_1$, $p_2^*=\ln0.3$ & 0.061199 \\
$M_{12}\to M_0$ & $X_1$, then $X_2$ & $F_0$ & 0.069752 \\
$M_{13}\to M_1$ & $X_3$ & $F_1$, $p_3^*=\ln0.6$ & 0.052301 \\
$M_{13}\to M_0$ & $X_1$, then $X_3$ & $F_0$ & 0.197408 \\
$M_{123}\to M_{12}$ & $X_3$ & $F_{12}^{(3)}$ & 0.045050 \\
$M_{123}\to M_{13}$ & $X_2$ & $F_{13}^{(2)}$ & 0.002352 \\
$M_{123}\to M_1$ & $X_2$ and $X_3$ & $F_1^{(23)}$ & 0.054646 \\
$M_{123}\to M_0$ & $X_1$, then $X_2$ and $X_3$ & $F_0$ & 0.092146 \\
\bottomrule
\end{tabular}
\end{table}

Four features of Table~\ref{tab:actions} stand out.

\begin{enumerate}
\item \textbf{The basal species alone is by far the most persistent state.} The barrier $S_{1\to0}=0.430$ exceeds every other action by at least a factor of two, so for large $\Omega$ the mean time to total extinction is dominated by the residence time $T_1=\pi_1/r_{1\to0}$ in \eqref{eq59}.
\item \textbf{Coexistence is lost through the intermediate species.} The action for losing $X_2$ from coexistence, $0.00235$, is nineteen times smaller than that of the next exit, the loss of $X_3$ ($0.0451$). This reflects how close $X_2$ is to its feasibility boundary: $x_2^*=0.1$ and, by Lemma~\ref{lem:establishment}, its passive eigenvalue at $M_{13}$ is only $f_2^{(13)}=0.02$. As $x_2^*\to0$ the fixed points $M_{123}$ and $M_{13}$ merge and the barrier vanishes. The WKB rate for this transition is accurate only when $\Omega S\gg1$, i.e.\ $\Omega\gg425$. For smaller communities it must be supplemented by the near-bifurcation treatment of \citet{AssafMeerson2017}.
\item \textbf{Joint and total losses are disfavoured, and the joint-loss path is in effect sequential.} The direct exits $M_{123}\to M_1$ ($0.0546$) and $M_{123}\to M_0$ ($0.0921$) have larger actions than either single-species loss. Moreover, the optimal path for $M_{123}\to M_1$ does not cross phase space directly. It first approaches $F_{13}^{(2)}$, winds around it, and only then continues to $F_1^{(23)}$. In this regime, the apparently simultaneous loss of both consumers is in fact the loss of $X_2$ followed shortly by the loss of $X_3$.
\item \textbf{The two-species states differ sharply.} From $M_{13}$, losing the apex consumer ($0.0523$) is far easier than collapse through the basal species ($0.1974$). From $M_{12}$ the two exits are comparable ($0.0612$ and $0.0698$), so both routes contribute appreciably at moderate $\Omega$.
\end{enumerate}

\begin{figure}[p]
\centering
\includegraphics[width=0.72\textwidth]{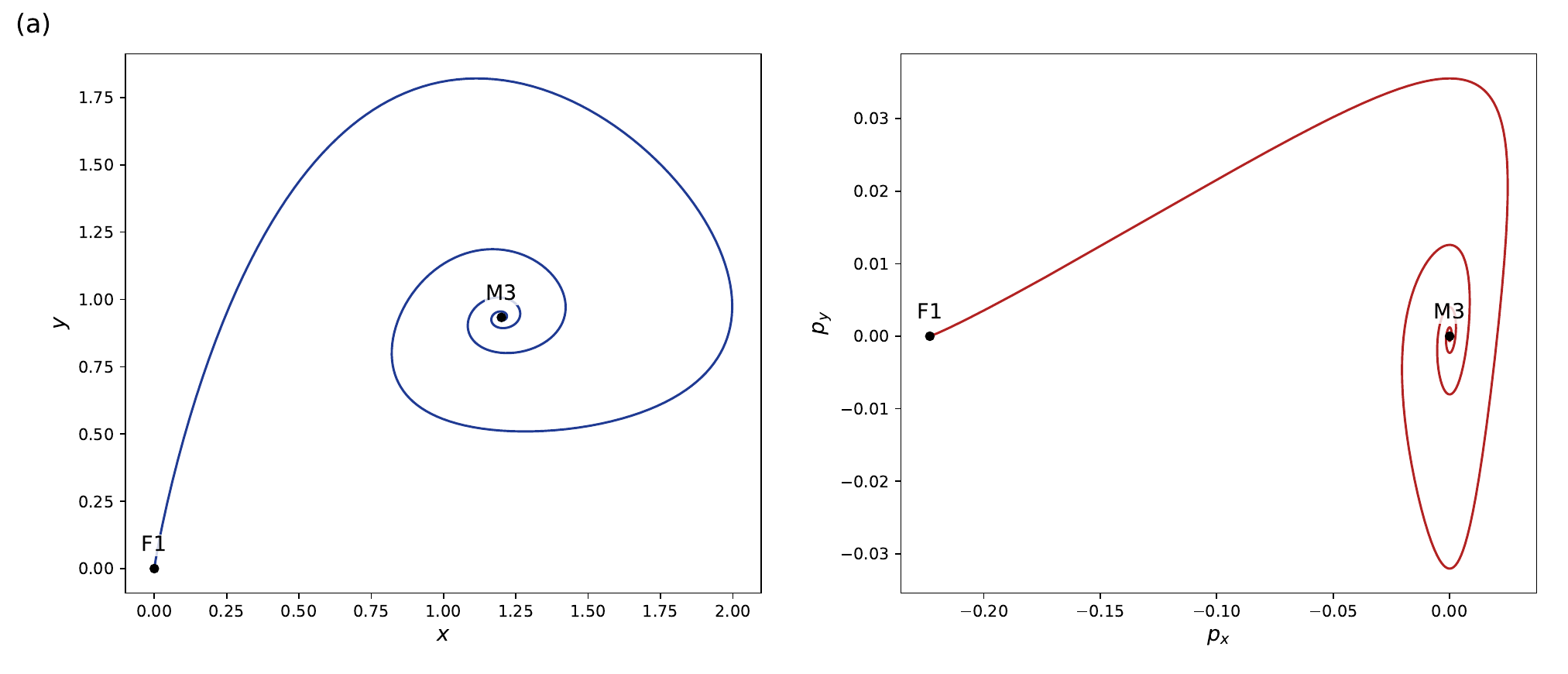}\\[2pt]
\includegraphics[width=0.72\textwidth]{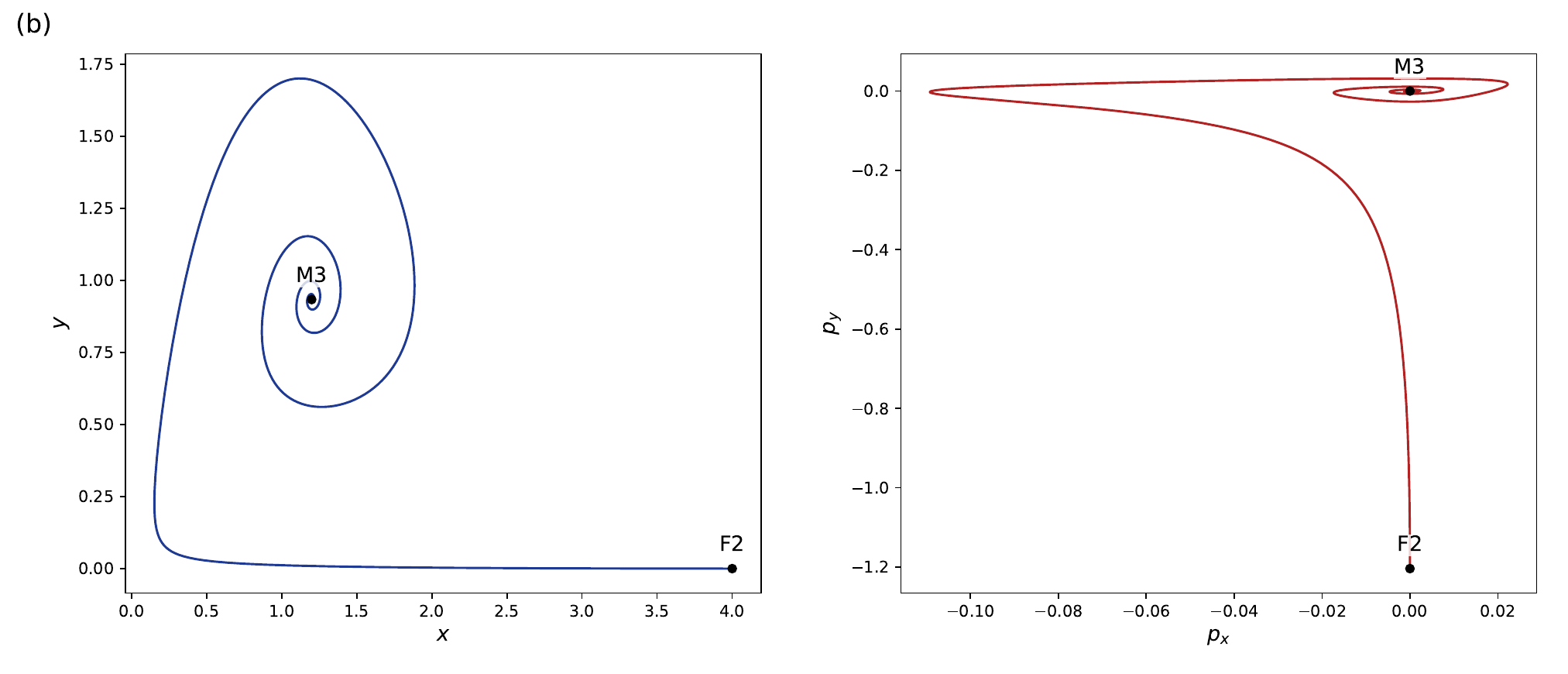}
\caption{Optimal extinction paths from the two-species state $M_{12}$, for the parameters of Section~\ref{sec:numerics}, in position space (left) and momentum space (right). (a) $M_{12}\to M_0$; (b) $M_{12}\to M_1$. In the panels, $M_3$, $F_1$ and $F_2$ denote $M_{12}$, $F_0$ and $F_1$ respectively}
\label{fig:instantons}
\vspace{10pt}
\includegraphics[width=0.72\textwidth]{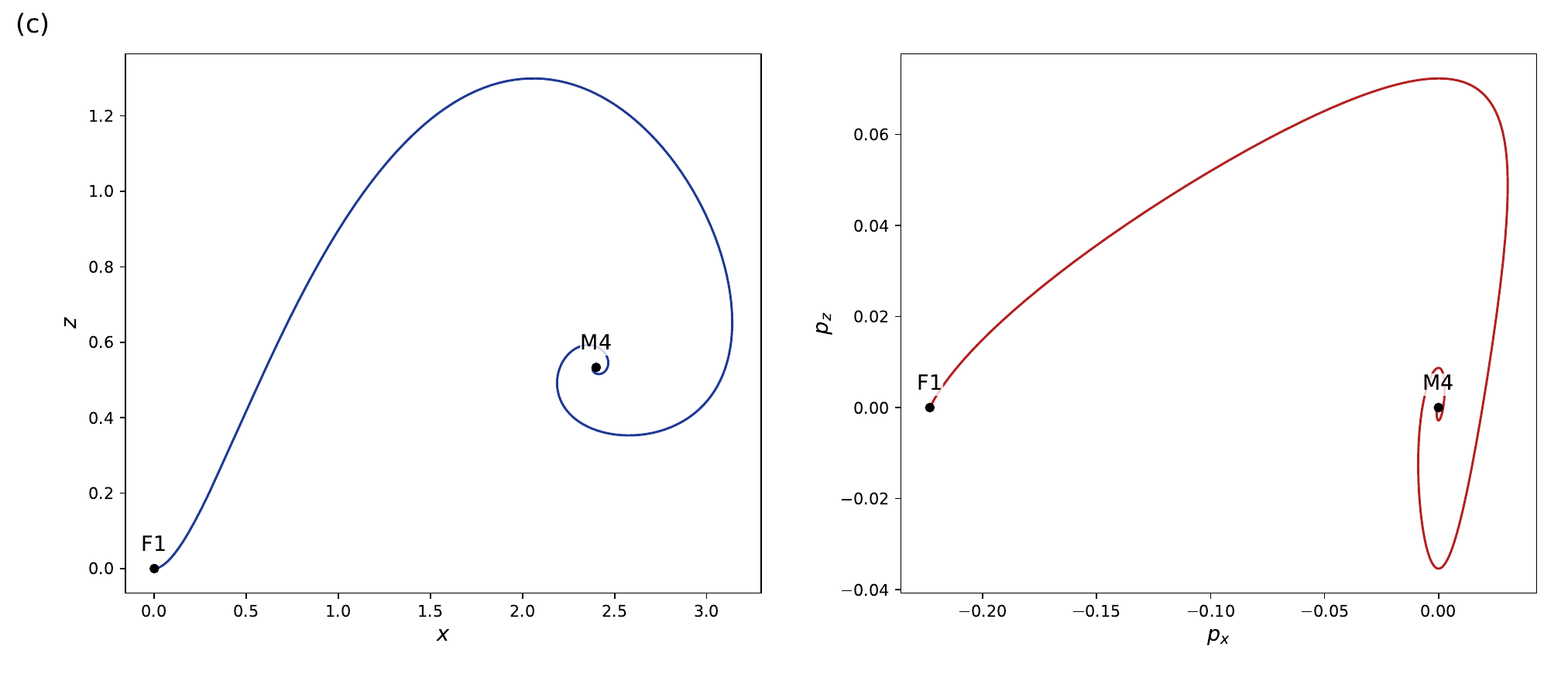}\\[2pt]
\includegraphics[width=0.72\textwidth]{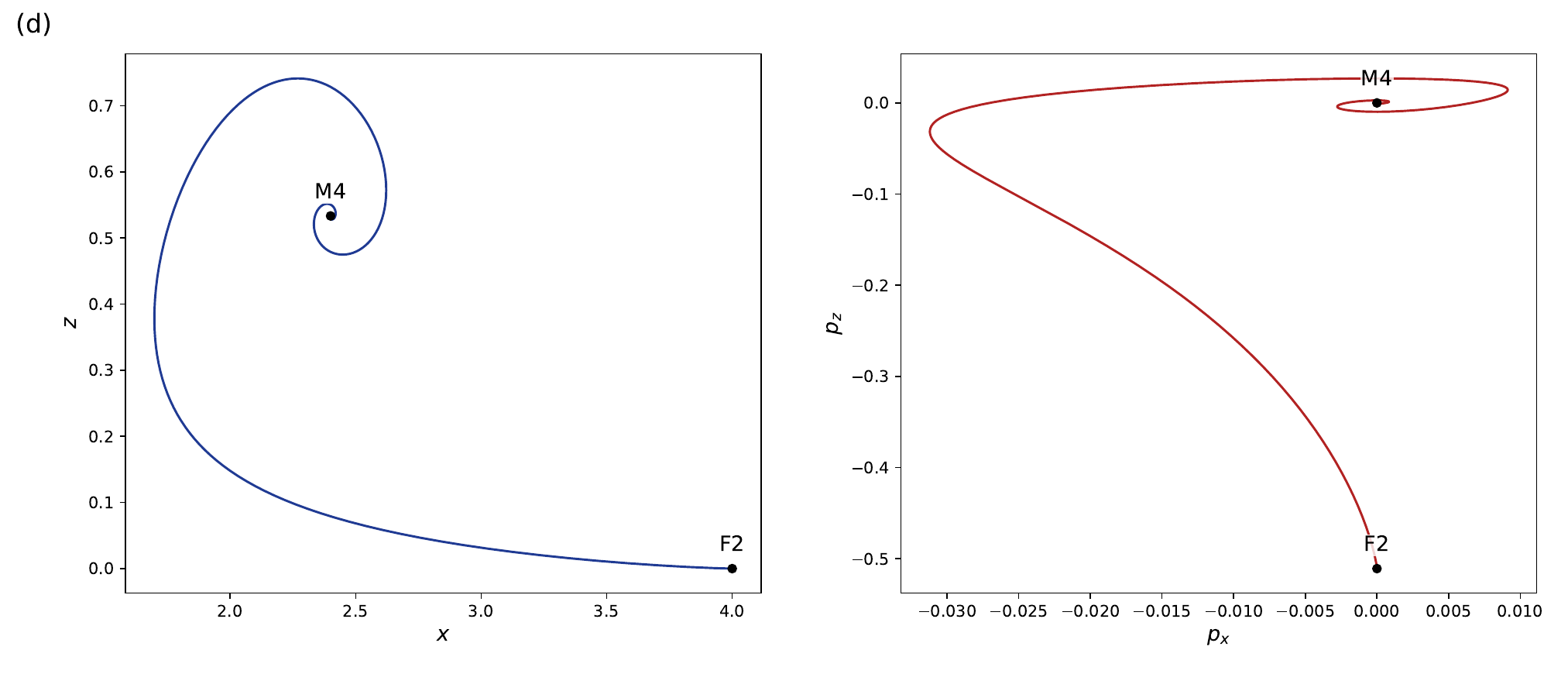}
\caption{Optimal extinction paths from the two-species state $M_{13}$, for the parameters of Section~\ref{sec:numerics}, in position space (left) and momentum space (right). (c) $M_{13}\to M_0$; (d) $M_{13}\to M_1$. In the panels, $M_4$, $F_1$ and $F_2$ denote $M_{13}$, $F_0$ and $F_1$ respectively}
\label{fig:instantons13}
\end{figure}

\begin{figure}[p]
\centering
\includegraphics[width=0.76\textwidth]{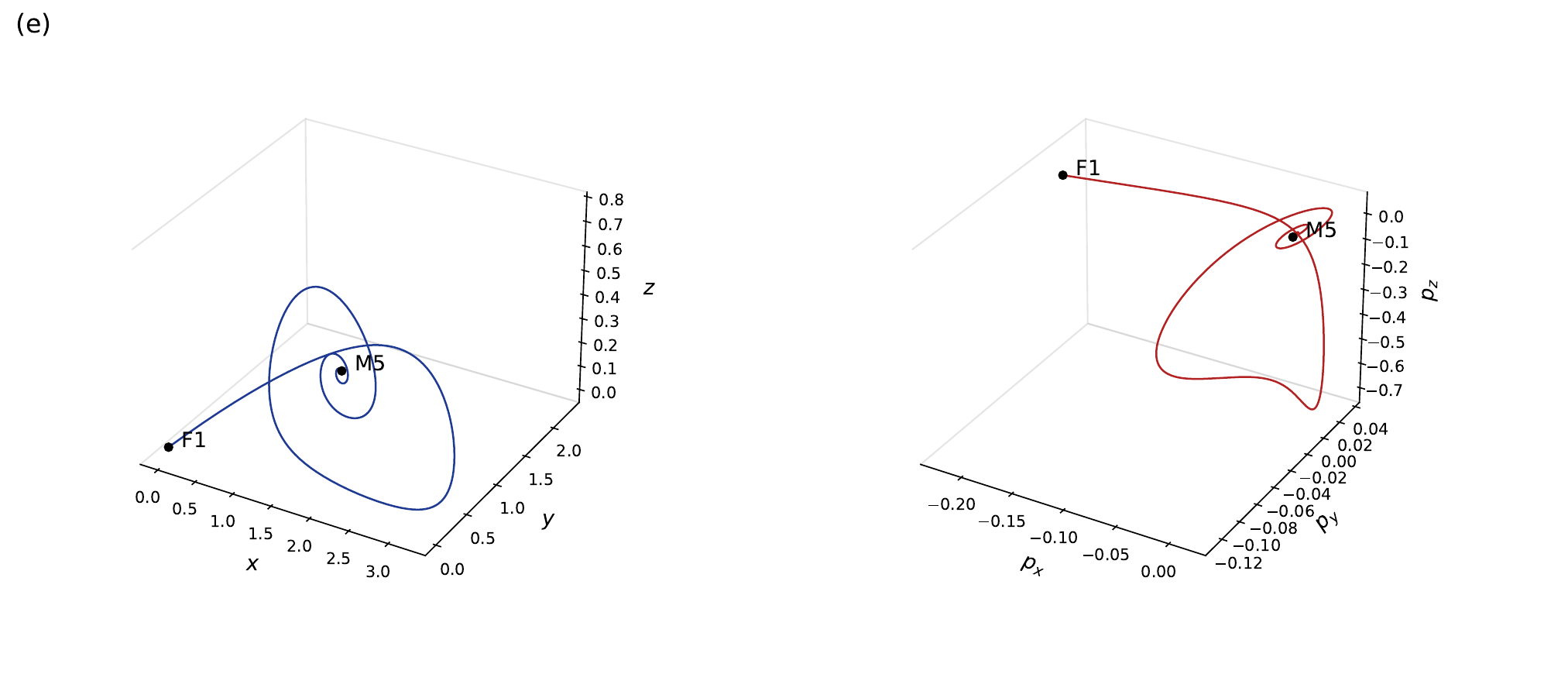}\\[2pt]
\includegraphics[width=0.76\textwidth]{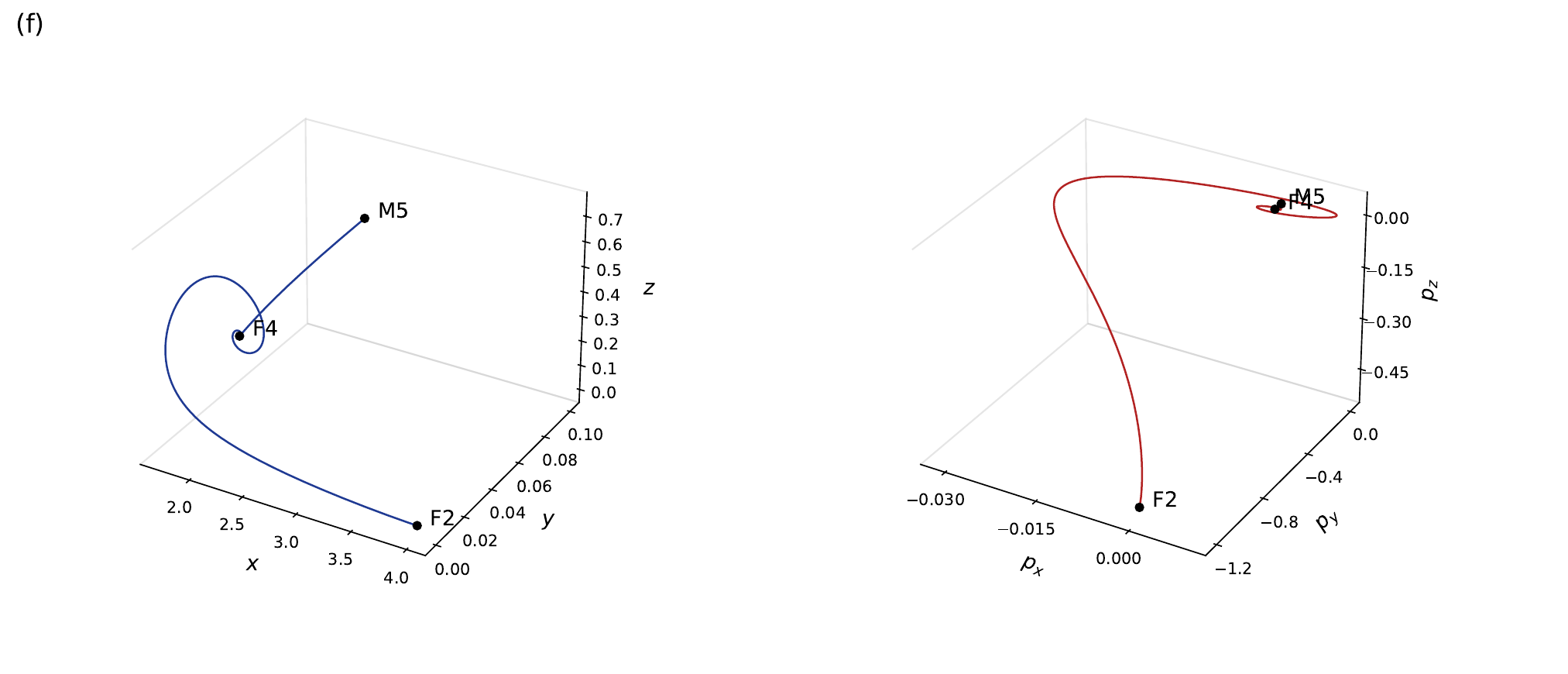}\\[2pt]
\includegraphics[width=0.76\textwidth]{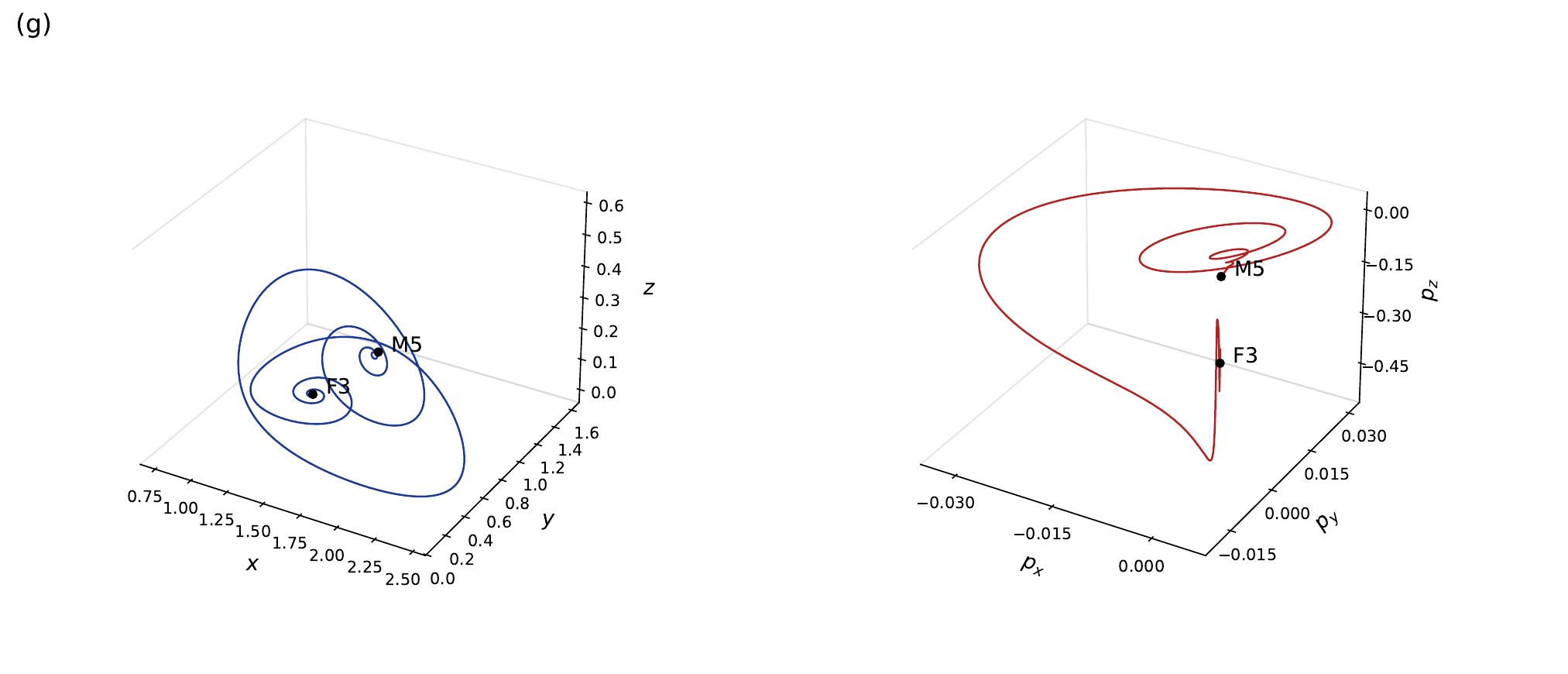}\\[2pt]
\includegraphics[width=0.76\textwidth]{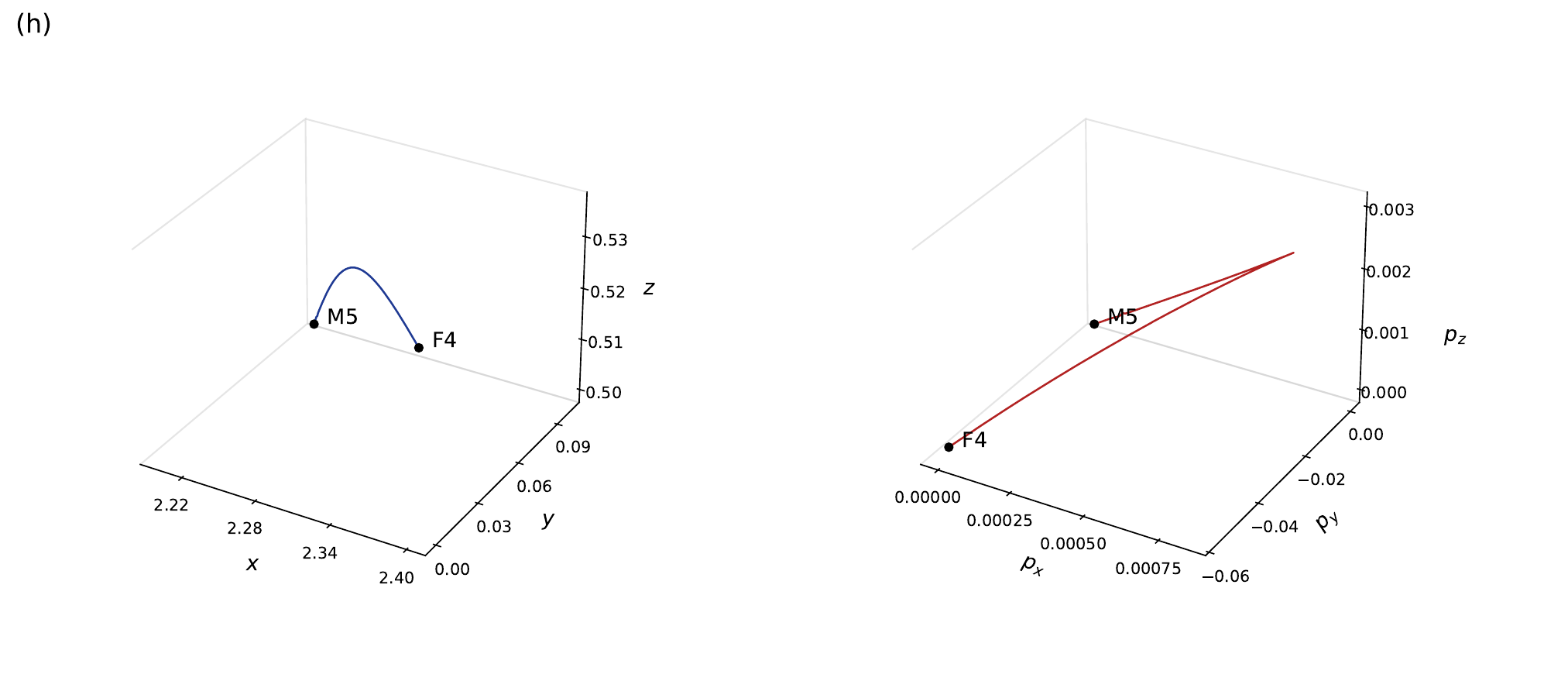}
\caption{Optimal extinction paths from the coexistence state $M_{123}$, for the parameters of Section~\ref{sec:numerics}, in position space (left) and momentum space (right). (e) $M_{123}\to M_0$; (f) $M_{123}\to M_1$, which passes close to $F_{13}^{(2)}$ before reaching $F_1^{(23)}$; (g) $M_{123}\to M_{12}$; (h) $M_{123}\to M_{13}$. In the panels, $M_5$, $F_1$, $F_2$, $F_3$ and $F_4$ denote $M_{123}$, $F_0$, $F_1$, $F_{12}^{(3)}$ and $F_{13}^{(2)}$ respectively}
\label{fig:instantons3}
\end{figure}

\paragraph{Geometry of the optimal paths.} Because $M_{12}$, $M_{13}$ and $M_{123}$ are foci, every instanton leaves its metastable state along an outward spiral. The spiral is most pronounced for $M_{123}\to M_{13}$, whose complex-mode component completes 9.6 turns on leaving $M_{123}$ and 6.1 turns on approaching $F_{13}^{(2)}$. Its amplitude grows by a factor of 5.5 per turn on departure and shrinks by 10.7 per turn on arrival. Both factors equal $\exp(2\pi|\operatorname{Re}\lambda|/\operatorname{Im}\lambda)$ for the eigenvalues at $M_{123}$ and $M_{13}$, which provides an independent check of the numerical trajectories. Figures~\ref{fig:instantons}--\ref{fig:instantons3} show the optimal paths in position and momentum space.

\begin{figure}[!htbp]
\centering
\includegraphics[width=120mm]{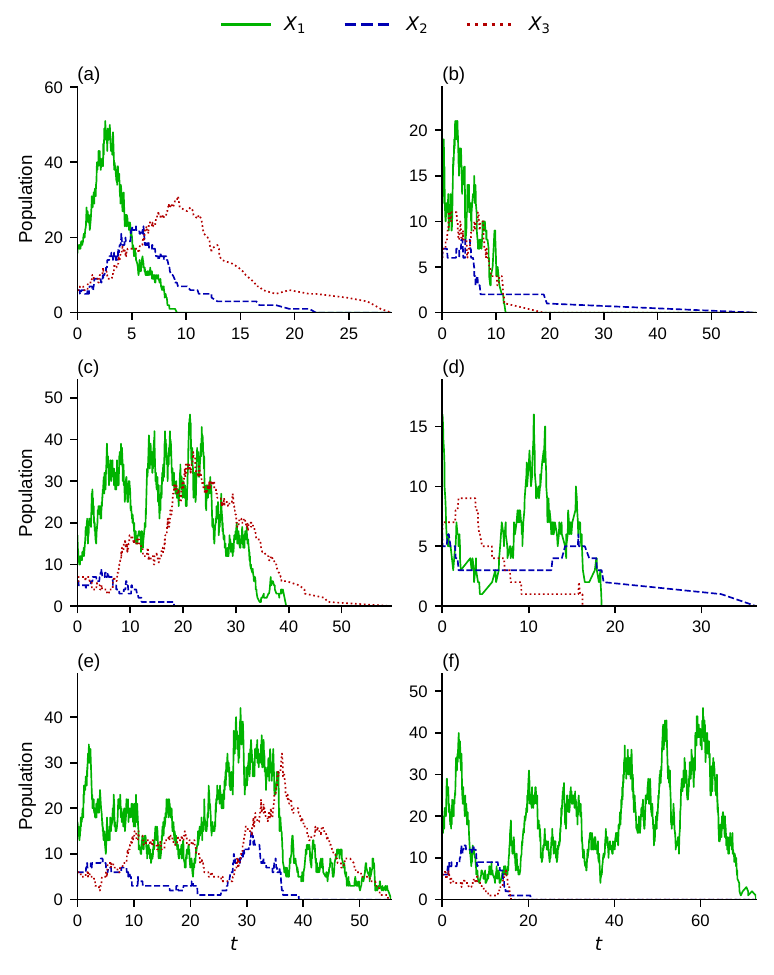}
\caption{Gillespie realisations of the three-species master equation, starting from the coexistence state. Populations of $X_1$ (solid), $X_2$ (dashed) and $X_3$ (dotted) are shown for one realisation of each of the six orders of extinction: (a) $X_1, X_2, X_3$; (b) $X_1, X_3, X_2$; (c) $X_2, X_1, X_3$; (d) $X_3, X_1, X_2$; (e) $X_2, X_3, X_1$; (f) $X_3, X_2, X_1$}
\label{fig:gillespie}
\end{figure}

\paragraph{Stochastic simulations.} Figure~\ref{fig:gillespie} shows Gillespie simulations of the master equation \eqref{eq16} at $\Omega=20$, for a second feasible parameter set ($b_1=1.0$, $d_1=0.85$, $a_{11}=0.056$, $d_2=0.056$, $d_3=0.213$, $a_{12}=0.201$, $a_{13}=0.136$, $a_{23}=0.332$, with coexistence at $(0.785,\,0.320,\,0.307)$). The six realisations exhibit all six orders in which the three species can disappear. In two, $X_1$ is lost first and the consumers then decline deterministically, as in the routes ending at $M_0$. In two, a single consumer is lost first, followed by collapse after $X_1$. In the remaining two, both consumers are lost and $X_1$ persists alone before going extinct last, as in the routes through $M_1$. At $\Omega=20$ the deterministic decline of the consumers after the loss of $X_1$ lasts tens of time units, which illustrates why the coarse-grained description requires $\Omega\gg1$.

\paragraph{Ordering of rates and the most probable extinction route.} Since $r\asymp\exp(-\Omega S)$, Table~\ref{tab:actions} orders the nine transition rates for $\Omega\gg1$ as
\begin{equation}\label{eq:rateorder}
r_{123\to13} \gg r_{123\to12} > r_{13\to1} > r_{123\to1} > r_{12\to1} > r_{12\to0} > r_{123\to0} > r_{13\to0} \gg r_{1\to0}.
\end{equation}
The branching ratios that make up \eqref{eq:pathsum} follow from this ordering. From $M_{123}$ the loss of $X_2$ is favoured over every other exit by a factor of at least $\exp[\Omega(S_{123\to12}-S_{123\to13})]=\exp(0.0427\,\Omega)$. From $M_{13}$ the loss of $X_3$ is favoured over collapse by $\exp(0.145\,\Omega)$, and $M_1$ has the single exit $M_1\to M_0$. The most probable extinction route is therefore
\begin{equation}\label{eq:mpp}
M_{123}\;\xrightarrow{\;X_2\;}\;M_{13}\;\xrightarrow{\;X_3\;}\;M_1\;\xrightarrow{\;X_1\;}\;M_0 ,
\end{equation}
in which the intermediate species is lost first, the apex consumer second and the basal species last, and its probability tends to one as $\Omega\to\infty$. Gillespie simulations of the master equation \eqref{eq16} for the same parameters, each started at coexistence and followed until $M_1$ is reached or $X_1$ is lost, confirm this route. It is followed by 83\% of 400 histories at $\Omega=40$ and by 85\% at $\Omega=80$. The time spent along the route is very unequal. The first step is fast, the second is slower, and the mean time to total extinction is dominated by the residence time $T_1$ of the basal species alone, as anticipated in Section~\ref{sec:three-rates}.

\begin{figure}[ht]
\centering
\includegraphics[width=\textwidth]{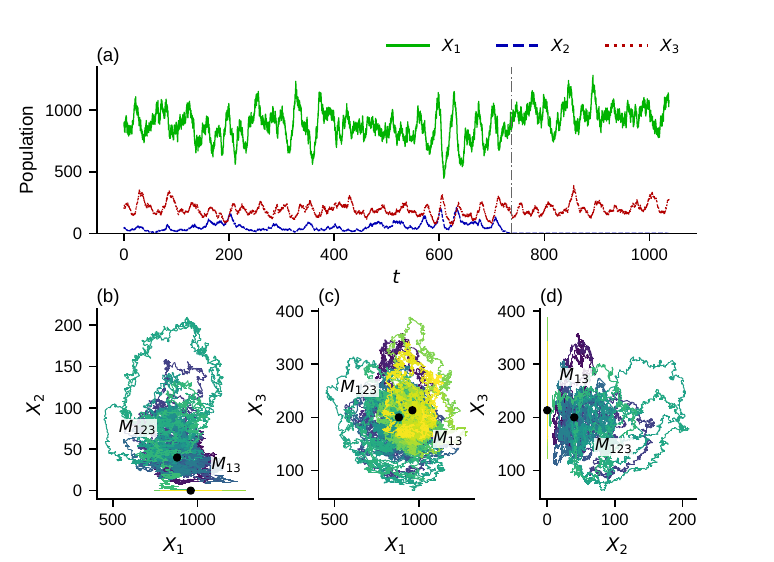}
\caption{A Gillespie realisation at $\Omega=400$ for the parameters of Table~\ref{tab:actions}. (a)~Populations of $X_1$ (solid), $X_2$ (dashed) and $X_3$ (dotted) against time, where the dash-dotted vertical line marks the extinction of $X_2$. (b)--(d)~Projections of the same trajectory on the $X_1$--$X_2$, $X_1$--$X_3$ and $X_2$--$X_3$ planes, shaded from dark (early) to light (late), with $\Omega M_{123}$ and $\Omega M_{13}$ marked}
\label{fig:case_graph}
\end{figure}

\paragraph{A single realisation at large $\Omega$.} Figure~\ref{fig:case_graph} follows one realisation at $\Omega=400$, started at coexistence. For several hundred time units the three populations fluctuate about $\Omega M_{123}$, and the projections show the noise-sustained spiral motion expected near a stable focus. The intermediate species, whose population remains small throughout, is lost at $t\approx737$. The trajectory then relaxes onto the plane $X_2=0$ and continues to circulate about $\Omega M_{13}$. This is the first step of the route \eqref{eq:mpp}. The subsequent loss of $X_3$ is exponentially rarer in $\Omega$ and is not observed on this time scale.

\subsection{Reduced topologies}\label{sec:reducedtop}

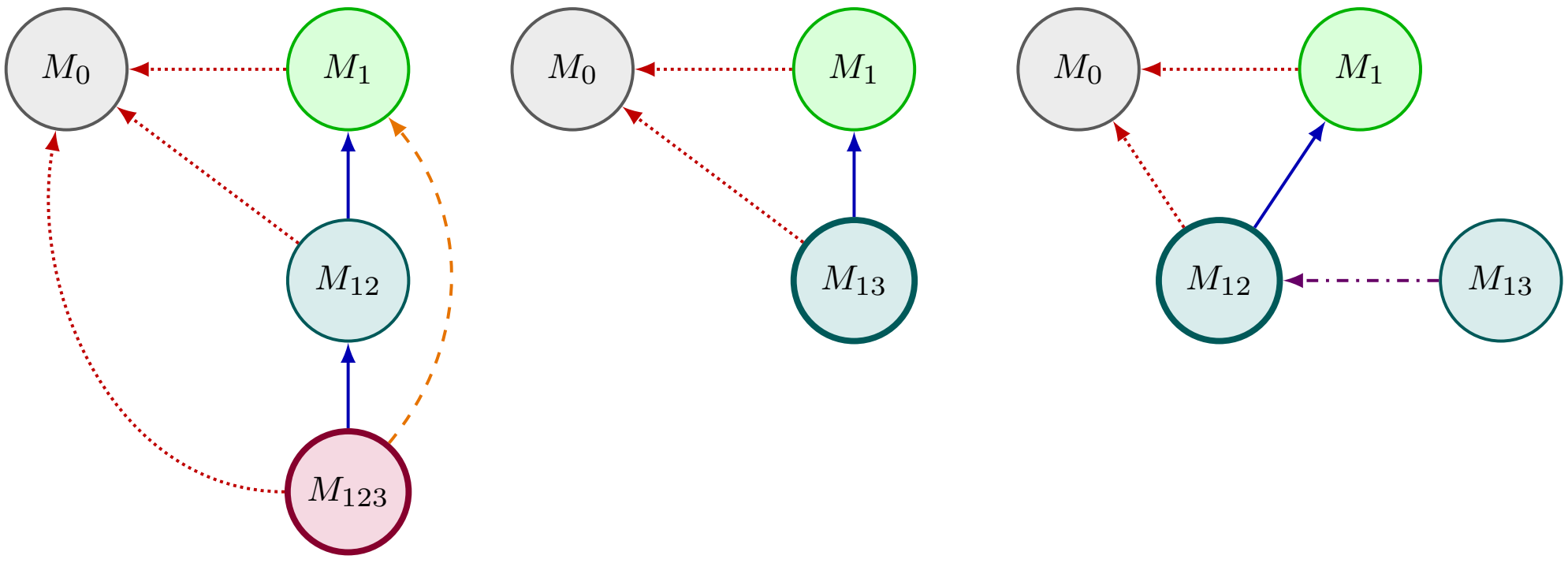
\begin{figure}[ht]
\centering
\resizebox{\textwidth}{!}{%
\begin{tikzpicture}[x=1cm,y=1cm,
  st/.style={circle,draw=#1!70!black,fill=#1!15,thick,minimum size=11mm,inner sep=0pt,font=\small},
  att/.style={line width=1.6pt},
  ed/.style={-latex,thick,blue!70!black},
  edj/.style={-latex,thick,dashed,orange!90!black},
  ed0/.style={-latex,thick,densely dotted,red!75!black},
  inv/.style={-latex,thick,dash dot,violet!80!black}]
\begin{scope}
\node[st=gray] (0) at (0,4.8) {$M_0$};
\node[st=green] (1) at (2.56,4.8) {$M_1$};
\node[st=teal] (12) at (2.56,2.88) {$M_{12}$};
\node[st=purple,att] (123) at (2.56,0.96) {$M_{123}$};
\draw[ed0] (1) -- (0);
\draw[ed] (12) -- (1);
\draw[ed0] (12) -- (0);
\draw[ed] (123) -- (12);
\draw[edj] (123) to[bend right=40] (1);
\draw[ed0] (123) to[out=180,in=-100] (0);
\node at (1.28,0.0) {(a) $a_{13}=0$};
\end{scope}
\begin{scope}[xshift=4.6cm]
\node[st=gray] (0) at (0,4.8) {$M_0$};
\node[st=green] (1) at (2.56,4.8) {$M_1$};
\node[st=teal,att] (13) at (2.56,2.88) {$M_{13}$};
\draw[ed0] (1) -- (0);
\draw[ed] (13) -- (1);
\draw[ed0] (13) -- (0);
\node at (1.28,0.0) {(b) $a_{12}=0$};
\end{scope}
\begin{scope}[xshift=9.2cm]
\node[st=gray] (0) at (0,4.8) {$M_0$};
\node[st=green] (1) at (2.56,4.8) {$M_1$};
\node[st=teal,att] (12) at (1.28,2.88) {$M_{12}$};
\node[st=teal] (13) at (3.84,2.88) {$M_{13}$};
\draw[ed0] (1) -- (0);
\draw[ed] (12) -- (1);
\draw[ed0] (12) -- (0);
\draw[inv] (13) -- node[below,font=\scriptsize]{} (12);
\node at (1.92,0.0) {(c) $a_{23}=0$, $R_2^{*}<R_3^{*}$};
\end{scope}
\end{tikzpicture}%
}
\caption{Extinction networks of the reduced topologies: (a) the food chain, $a_{13}=0$ (b) the intermediate species without prey, $a_{12}=0$ (c) exploitative competition, $a_{23}=0$, shown for $R_2^*<R_3^*$. Line styles follow Fig.~\ref{fig:network}: single-species losses are solid, the simultaneous loss of both consumers is dashed and losses of $X_1$ are dotted. Thick borders mark the deterministic attractors, and the dash-dotted arrow in (c) is the deterministic establishment of $X_2$ at $M_{13}$}
\label{fig:reduced}
\end{figure}

Deleting one of the three trophic links of the omnivory module yields the other three hierarchical topologies. Each deletion removes some fixed points, changes the stability of others, and prunes the extinction network (Fig.~\ref{fig:reduced}). The results are summarised in Table~\ref{tab:reduced} and derived below. Let $R_k^*\equiv d_k/a_{1k}$ denote the density of $X_1$ at which consumer $X_k$, feeding on $X_1$ alone, exactly balances its death rate \citep{Tilman1982}.

\begin{table}[ht]
\centering
\caption{Effect of deleting a single trophic link}
\label{tab:reduced}
\small
\setlength{\tabcolsep}{4pt}
\begin{tabular}{>{\raggedright}p{1.5cm}>{\raggedright}p{3.0cm}>{\raggedright}p{2.9cm}>{\raggedright}p{2.3cm}>{\raggedright\arraybackslash}p{1.7cm}}
\toprule
Deleted link & Topology & Feasible states & Attractor & Transitions \\
\midrule
none & omnivory module & $M_0$, $M_1$, $M_{12}$, $M_{13}$, $M_{123}$ & $M_{123}$ & 9 \\
$a_{13}=0$ & food chain $X_1\to X_2\to X_3$ & $M_0$, $M_1$, $M_{12}$, $M_{123}$ & $M_{123}$ & 6 \\
$a_{12}=0$ & $X_3$ eats $X_1$ and $X_2$ & $M_0$, $M_1$, $M_{13}$ & $M_{13}$ & 3 \\
$a_{23}=0$ & $X_2$ and $X_3$ compete for $X_1$ & $M_0$, $M_1$, $M_{12}$, $M_{13}$ & state of the consumer with smaller $R^*$ & 3 \\
\bottomrule
\end{tabular}
\end{table}

\paragraph{Case 1: $a_{13}=0$ (food chain).} Without the omnivorous link, $M_{13}$ ceases to exist, while $M_0$, $M_1$ and $M_{12}$ are unchanged. Setting $a_{13}=0$ in \eqref{eq38} gives
\begin{equation}\label{eq64}
M_{123} = \Big(\frac{a_{23}B_1 - a_{12}d_3}{a_{11}a_{23}},\ \frac{d_3}{a_{23}},\ \frac{a_{23}u - a_{12}^2d_3}{a_{11}a_{23}^2}\Big), \qquad \text{feasible iff}\quad a_{23}\,u > a_{12}^2\,d_3.
\end{equation}
This single inequality implies $u>0$ and $x_1^*>0$, so it guarantees all four states. At $M_1$ the eigenvalues are $-B_1$, $u/a_{11}>0$ and $-d_3<0$, since the apex species cannot feed on $X_1$ alone, so $M_1$ is a saddle with one unstable direction. $M_{12}$ remains a saddle by \eqref{eq44}. At $M_{123}$, \eqref{eq47} gives $c_1c_2-c_0=a_{11}a_{12}^2x_1^{*2}x_2^*>0$, and the argument of Proposition~\ref{prop:global} carries over, so coexistence is globally stable. The network keeps six transitions: $r_{1\to0}$, $r_{12\to1}$, $r_{12\to0}$, $r_{123\to12}$, $r_{123\to1}$ and $r_{123\to0}$. At $M_1$ the apex species has no prey, $G_3=0$, so its momentum has only the root zero and only $F_1^{(2)}$ survives there. Consequently the instanton for $M_{123}\to M_1$ ends at $F_1^{(2)}$. The remaining fluctuational point at a two-species state is
\begin{equation}\label{eq65}
F_{12}^{(3)}:\ \ p_3^{*} = \ln\frac{a_{12}^2\,d_3}{a_{23}\,u}.
\end{equation}

\paragraph{Case 2: $a_{12}=0$ (intermediate species without prey).} $X_2$ now has no food source and acts as a pure sink. $M_{12}$ ceases to exist, and $M_{123}$ is never feasible, since \eqref{eq38} gives $x_3^*=-d_2/a_{23}<0$. The remaining states $M_0$, $M_1$ and $M_{13}$ all exist if and only if $v>0$. At $M_1$ the eigenvalues are $-B_1$, $-d_2<0$ and $v/a_{11}>0$, so $M_1$ is a saddle. At $M_{13}$ the passive eigenvalue of $X_2$ becomes $f_2=-(d_2+a_{23}x_3)<0$ and the active block is stable by \eqref{eq45}, so $M_{13}$ is asymptotically stable and is the attractor of the system. The dynamics reduce to a two-species predator-prey system with self-limited prey, of the type studied in \citet{GottesmanMeerson2012}, with three transitions $r_{1\to0}$, $r_{13\to1}$ and $r_{13\to0}$. Because $X_2$ has no prey, its momentum admits only the root zero, and the fluctuational points are $F_0$ and $F_1^{(3)}$.

\paragraph{Case 3: $a_{23}=0$ (exploitative competition).} The two consumers now share the resource $X_1$ without interacting directly. $M_{12}$ and $M_{13}$ exist if $u>0$ and $v>0$, respectively. Coexistence is impossible, because $\det \mathbf L_{\{123\}}=a_{11}a_{23}^2=0$, and the two consumer equations would require $x_1=R_2^*$ and $x_1=R_3^*$ simultaneously. The passive eigenvalues are
\begin{equation}\label{eq66}
f_3^{(12)} = a_{13}\big(R_2^{*} - R_3^{*}\big), \qquad f_2^{(13)} = a_{12}\big(R_3^{*} - R_2^{*}\big),
\end{equation}
which have opposite signs. If $R_2^*<R_3^*$, then $M_{12}$ is asymptotically stable (its active block is stable by \eqref{eq45}) and $M_{13}$ is a saddle at which $X_2$ can establish itself. If $R_2^*>R_3^*$, the roles are reversed. The consumer that can persist at the lower resource density excludes the other, which is the $R^*$ rule of resource competition theory \citep{Tilman1982}. Starting with both consumers present, the deterministic flow carries the system to the winner's two-species state, from which the network has three transitions, for example $r_{12\to1}$, $r_{12\to0}$ and $r_{1\to0}$ when $X_2$ wins. The losing state is reached only if the winner is not present initially. The fluctuational momenta at the two-species states,
\begin{equation}\label{eq67}
F_{12}^{(3)}:\ \ p_3^{*} = \ln\frac{R_3^{*}}{R_2^{*}}, \qquad F_{13}^{(2)}:\ \ p_2^{*} = \ln\frac{R_2^{*}}{R_3^{*}},
\end{equation}
are equal in magnitude and opposite in sign, mirroring the opposite signs in \eqref{eq66}. The momentum is positive at the winner's state, where the loser cannot establish, and negative at the loser's state.

\section{Conclusions}\label{sec:conclusions}

We have developed a WKB theory of demographic extinction for hierarchical food webs of arbitrary size and solved the three-species omnivory module explicitly as its first non-trivial case.

Three structural facts make the general problem tractable. First, the feasible states are labelled by active sets containing the basal species, and the determinants of their defining equations have signs that alternate with community size (Lemma~\ref{lem:det}). Second, this alternation turns establishment analysis into a sign rule: an extinct species can establish exactly when its density at the enlarged fixed point is positive (Lemma~\ref{lem:establishment}). Third, with unit conversion efficiency the interaction matrix is antisymmetric apart from basal self-limitation, which confines every resident eigenvalue to the closed left half-plane (Lemma~\ref{lem:active}) and supplies a Volterra Lyapunov function. Together these imply that every partial community is a saddle whose unstable directions all point out of its own extinction hyperplane. Each partial community is therefore metastable once the corresponding species have been lost, and full coexistence is the unique deterministic attractor.

On the stochastic side, the WKB Hamiltonian decouples the momenta of the extinct species at every partial community. The resulting fluctuational fixed points have the closed form $p_k=\ln(D_k/G_k)$, the logarithm of the ratio of loss to gain rates of a rare entrant. The extinction network has $2^{N-1}+1$ nodes and $3^{N-1}$ edges, and its coarse-grained dynamics are triangular with respect to set inclusion. This yields an exact recursion for all mean residence and extinction times, and expresses the probability of every extinction route as a product of branching ratios.

For three species the theory gives three existence conditions of mutual establishment form, global stability of coexistence, the six fluctuational fixed points, and closed forms for the mean extinction times $\tau_2$, $\tau_3$, $\tau_{23}$ and $\tau_{\mathrm{tot}}$ in terms of nine transition rates. The numerical actions for one parameter set show a clear hierarchy of barriers. The basal species alone is far more persistent than any larger community, coexistence is lost first through the species closest to its feasibility boundary, and the apparently simultaneous loss of both consumers proceeds as a rapid succession. The reduced topologies show how strongly this structure depends on omnivory. Deleting the omnivorous link leaves a food chain with six transitions, deleting the only food source of the intermediate species collapses the system to a two-species predator-prey problem, and deleting the link between the consumers replaces coexistence by competitive exclusion under the $R^*$ rule.

Four problems remain open.
\begin{enumerate}
\item \textbf{Prefactors, simulations and parameter dependence.} The actions have been computed for one parameter set. Pre-exponential factors, a quantitative comparison of \eqref{eq:pathsum} and \eqref{eq30} with mean extinction times measured in Gillespie simulations over a range of $\Omega$, and a systematic exploration of parameter space would show whether the observed ordering of barriers holds beyond the parameter set studied.
\item \textbf{Minimal existence conditions.} A minimal independent set of feasibility inequalities is known only for $N=3$. Exact symbolic elimination, for instance by Gr\"obner bases, is a natural route for $N\ge4$.
\item \textbf{Beyond antisymmetry.} Relaxing unit conversion efficiency, or adding self-limitation at higher trophic levels, breaks the antisymmetry on which Lemma~\ref{lem:active} rests and may produce oscillatory, chaotic or multistable attractors. Extinction would then have to be treated from limit cycles or through switching between attractors.
\item \textbf{Alternative topologies.} The cyclic, non-hierarchical topology and food webs with several basal species lie outside the present framework.
\end{enumerate}

More broadly, the extinction of a structured community is not a single rare event but a cascade on a combinatorially rich network. The topology of that network is fixed by the deterministic structure, and its rates by instanton actions. The framework developed here provides the scaffolding on which those rates can be computed systematically for food webs of any size.

\begin{appendices}
\renewcommand{\theequation}{\thesection\arabic{equation}}
\section{Food webs as directed graphs}\label{app:graphs}
\setcounter{equation}{0}

A food web is formally a directed graph (digraph) in which each vertex represents a distinct biological taxon and each directed edge represents a trophic interaction, oriented here from prey to predator.

A labelled graph on $N$ vertices is one in which each vertex carries a distinct and fixed identity, corresponding in the ecological context to a distinguishable species. Two labelled graphs are considered identical only if they share both the same vertex set and the same edge set. An unlabelled graph, by contrast, treats all graphs related by a permutation of the vertices as equivalent, so that two graphs are considered the same if one can be obtained from the other by simply renaming the nodes. The count of unlabelled graphs enumerates abstract
structural patterns corresponding to the wiring architectures that recur across ecosystems regardless of taxonomic composition.

An oriented graph is a directed graph in which every pair of vertices is connected by at most one directed edge. The exclusion of bidirected edges rules out mutual predation, which is occasionally observed in intraguild predation systems, and is absent from the idealised hierarchical food web model adopted here. We count only connected graphs, i.e.\ networks in which every vertex is reachable from every other in the underlying undirected sense. This constraint is ecologically appropriate because a food web containing a fully disconnected subset of species would represent two independent communities rather than a single coherent web, and connectivity therefore serves as the minimum structural criterion for ecological coherence.

For labelled graphs the connected counts follow from the totals by the standard recursion
\begin{equation}\label{eq:connected}
c_N = t_N - \sum_{k=1}^{N-1}\binom{N-1}{k-1}\,c_k\,t_{N-k}, \qquad t_N = q^{\binom{N}{2}},
\end{equation}
which removes every graph whose vertex 1 lies in a component of size $k<N$. Here $q=2$ for simple graphs, in which each pair of vertices is either joined or not, and $q=3$ for oriented graphs, in which each pair is unjoined or joined in one of two directions. The three sequences of Table~\ref{tab:oeis} delineate the space of food-web architectures.

\paragraph{OEIS A001187: connected labelled undirected graphs.} A001187($N$) counts the connected simple graphs on $N$ labelled vertices, without any directionality constraint on edges. Once a strict trophic ranking of the $N$ labelled species is fixed, every edge acquires a unique orientation, from the lower-ranked species to the higher-ranked one, and the resulting digraph is acyclic. A001187($N$) therefore also counts the connected hierarchical food webs on $N$ species with a prescribed trophic ranking, each of which is a realisable food web in which species occupy distinct, ordered trophic positions. For $N=3$ its value, 4, is exactly the number of hierarchical topologies treated in Section~\ref{sec:three}: the omnivory module and the three webs obtained by deleting a single link (Section~\ref{sec:reducedtop}).

\paragraph{OEIS A086345: connected unlabelled oriented graphs.} A086345($N$) counts the connected oriented graphs on $N$ unlabelled vertices, i.e.\ isomorphism classes of connected directed graphs with no bidirected edges. This sequence enumerates the space of possible food-web topologies: the distinct structural archetypes, up to vertex relabelling. It includes webs that contain directed cycles, which no strict hierarchy can produce. For $N=3$ its five members are the four hierarchical topologies above together with the cyclic web $X_1\to X_2\to X_3\to X_1$. Two food webs from geographically separated ecosystems that belong to the same class are structurally equivalent regardless of their taxonomic composition, which makes this sequence the natural basis for cross-ecosystem structural comparison.

\paragraph{OEIS A054941: connected labelled oriented graphs.} A054941($N$) counts the connected oriented graphs on $N$ labelled vertices, i.e.\ directed graphs with no bidirected edges in which each vertex represents a distinct, identifiable species. This sequence counts the distinct realisable food webs for a community of $N$ biologically distinguishable species. A001187 is its hierarchical counterpart; together the two delineate the full combinatorial space of food-web architectures, with and without the imposition of a fixed trophic order.

\begin{table}[ht]
\centering
\caption{Numbers of connected food-web graphs on $N$ species}
\label{tab:oeis}
\small\setlength{\tabcolsep}{4pt}
\begin{tabular}{llrrrrrrr}
\toprule
Sequence & Description & $N=1$ & 2 & 3 & 4 & 5 & 6 & 7 \\
\midrule
A001187 & labelled, hierarchical & 1 & 1 & 4 & 38 & 728 & 26\,704 & 1\,866\,256 \\
A086345 & unlabelled, oriented & 1 & 1 & 5 & 34 & 535 & 20\,848 & 2\,120\,098 \\
A054941 & labelled, oriented & 1 & 2 & 20 & 624 & 55\,248 & 13\,982\,208 & 10\,358\,360\,640 \\
\bottomrule
\end{tabular}
\end{table}

The model of Section~\ref{sec:model} is the complete hierarchy, the densest member of A001187($N$). Its sub-webs, obtained by deleting predatory links, exhaust the remaining hierarchical topologies.

\section{Proofs}\label{app:proofs}
\setcounter{equation}{0}

\paragraph{B.1 Proof of Lemma~\ref{lem:det}.} Write $\mathbf L_A=\begin{pmatrix}a_{11}&\mathbf u^{\T}\\ \mathbf u&\boldsymbol\Lambda'_A\end{pmatrix}$, where $\boldsymbol\Lambda'_A$ is an antisymmetric matrix of order $m=n-1$. Expanding along the first row and column gives, for any square $\boldsymbol\Lambda'_A$,
\begin{equation}\label{eqB2}
\det \mathbf L_A = a_{11}\det\boldsymbol\Lambda'_A - \mathbf u^{\T}\operatorname{adj}(\boldsymbol\Lambda'_A)\,\mathbf u.
\end{equation}
Since $\operatorname{adj}(\boldsymbol\Lambda'_A)^{\T}=\operatorname{adj}(\boldsymbol\Lambda'^{\T}_A)=\operatorname{adj}(-\boldsymbol\Lambda'_A)=(-1)^{m-1}\operatorname{adj}(\boldsymbol\Lambda'_A)$, the adjugate is antisymmetric when $m$ is even and symmetric when $m$ is odd.

{$n$ odd ($m$ even).} The adjugate is antisymmetric, so $\mathbf u^{\T}\operatorname{adj}(\boldsymbol\Lambda'_A)\mathbf u=0$, and $\det\boldsymbol\Lambda'_A=\operatorname{Pf}(\boldsymbol\Lambda'_A)^2$. Hence $\det \mathbf L_A=a_{11}\operatorname{Pf}(\boldsymbol\Lambda'_A)^2\ge0$.

{$n$ even ($m$ odd).} An antisymmetric matrix of odd order is singular, so $\det\boldsymbol\Lambda'_A=0$ and its rank is at most $m-1$, so the adjugate has rank at most one. It is symmetric, and its diagonal entries are the determinants of the antisymmetric principal minors of even order $m-1$, which are squares of Pfaffians and hence non-negative. A symmetric matrix of rank at most one with non-negative diagonal has the form $\mathbf w\mathbf w^{\T}$, so $\det \mathbf L_A=-(\mathbf u^{\T}\mathbf w)^2\le0$. Moreover $\boldsymbol\Lambda'_A\operatorname{adj}(\boldsymbol\Lambda'_A)=\det(\boldsymbol\Lambda'_A)\,\mathbf I=0$, so $\mathbf w$ spans the kernel of $\boldsymbol\Lambda'_A$ whenever $\mathbf w\neq0$. \hfill$\square$

\paragraph{B.2 Block-triangular spectrum.} Listing the active species of $M_A$ first is a simultaneous permutation of rows and columns, which is a similarity transformation and preserves the spectrum. For the resulting block upper-triangular matrix \eqref{eq12},
\begin{equation}\label{eqB1}
\det\big(\mathbf J_A - \lambda \mathbf I\big) = \det\big(\mathbf J_A^{S} - \lambda \mathbf I\big)\,\det\big(\mathbf D_A - \lambda \mathbf I\big).
\end{equation}
In the Leibniz expansion of the left-hand side, any permutation that sends a row of the lower block to a column of the upper block picks up a factor from the zero block and vanishes. The surviving permutations map each block to itself, and the sum factorises into the two determinants on the right.

\paragraph{B.3 Proof of Lemma~\ref{lem:establishment}.} Order $A'=A\cup\{k\}$ with $k$ last; this simultaneous permutation leaves determinants and Cramer's rule unchanged, keeps species 1 first, and preserves the antisymmetry used in Lemma~\ref{lem:det}. Then
\begin{equation}\label{eqB3}
\mathbf L_{A'} = \begin{pmatrix} \mathbf L_A & \boldsymbol\ell \\ \mathbf m^{\T} & 0 \end{pmatrix}, \qquad \mathbf c_{A'} = \begin{pmatrix} \mathbf c_A \\ d_k \end{pmatrix},
\end{equation}
where the last row is the equilibrium condition of $X_k$, $\mathbf m^{\T}\mathbf x=d_k$, with $m_i=a_{ik}$ for $i<k$ and $m_j=-a_{kj}$ for $j>k$. Comparing with \eqref{eq13}, the passive eigenvalue at $M_A$ is $f_k^{(A)}=\mathbf m^{\T}\mathbf x^A-d_k$, where $\mathbf x^A=\mathbf L_A^{-1}\mathbf c_A$. By Cramer's rule and the Schur-complement formula for the determinant,
\begin{equation}\label{eqB4}
\begin{aligned} x_k^{(A')} &= \frac{1}{\det \mathbf L_{A'}}\det\begin{pmatrix} \mathbf L_A & \mathbf c_A \\ \mathbf m^{\T} & d_k \end{pmatrix} \\ &= \frac{\det \mathbf L_A}{\det \mathbf L_{A'}}\big(d_k - \mathbf m^{\T}\mathbf L_A^{-1}\mathbf c_A\big) = -\,\frac{\det \mathbf L_A}{\det \mathbf L_{A'}}\,f_k^{(A)}, \end{aligned}
\end{equation}
which is \eqref{eq14}. By Lemma~\ref{lem:det}, $\det \mathbf L_A$ and $\det \mathbf L_{A'}$ have the signs $(-1)^{n+1}$ and $(-1)^n$, so the ratio $-\det \mathbf L_{A'}/\det \mathbf L_A$ is positive. \hfill$\square$

\paragraph{B.4 Proof of Lemma~\ref{lem:active}.} At a feasible $M_A$ the matrix $\mathbf X_A=\operatorname{diag}(x_s)$ is positive definite, so $\mathbf J_A^S=\mathbf X_A\mathbf K_A$ is similar to
\begin{equation}\label{eqB5}
\mathbf M = \mathbf X_A^{1/2}\mathbf K_A\mathbf X_A^{1/2} = \mathbf X_A^{1/2}\boldsymbol\Lambda_A\mathbf X_A^{1/2} - a_{11}x_1\,\mathbf e_1\mathbf e_1^{\T},
\end{equation}
where $\boldsymbol\Lambda_A$, the restriction of $\boldsymbol\Lambda$ to the active species, is antisymmetric. For an eigenpair $\mathbf M\boldsymbol\theta=\lambda\boldsymbol\theta$ with complex eigenvector $\boldsymbol\theta$,
\begin{equation}\label{eqB6}
\begin{gathered}\lambda\,\lVert\boldsymbol\theta\rVert^2 = \boldsymbol\theta^{\dagger}\mathbf M\boldsymbol\theta = \underbrace{\boldsymbol\theta^{\dagger}\mathbf X_A^{1/2}\boldsymbol\Lambda_A\mathbf X_A^{1/2}\boldsymbol\theta}_{\text{purely imaginary}} - a_{11}x_1\,|\theta_1|^2, \\ \operatorname{Re}\lambda = -\,\frac{a_{11}x_1\,|\theta_1|^2}{\lVert\boldsymbol\theta\rVert^2} \le 0. \end{gathered}
\end{equation}
The first term is purely imaginary because $\mathbf X_A^{1/2}\boldsymbol\Lambda_A\mathbf X_A^{1/2}$ is real and antisymmetric. \hfill$\square$

\paragraph{B.5 Derivative of the Lyapunov function (Proposition~\ref{prop:lyapunov}).} At the coexistence state every $g_k$ vanishes, so $\mathbf g(\mathbf x)=\mathbf K(\mathbf x-\mathbf x^*)$. Then
\[
\dot V=\sum_k\Big(1-\frac{x_k^*}{x_k}\Big)\dot x_k=\sum_k(x_k-x_k^*)\,g_k(\mathbf x)=(\mathbf x-\mathbf x^*)^{\T}\mathbf K(\mathbf x-\mathbf x^*).
\]
Writing $\mathbf K=\boldsymbol\Lambda-a_{11}\mathbf e_1\mathbf e_1^{\T}$, the antisymmetric part drops out because $\mathbf u^{\T}\boldsymbol\Lambda\mathbf u=0$ for every $\mathbf u$, leaving $\dot V=-a_{11}(x_1-x_1^*)^2$.

\section{Stability criteria in three dimensions}\label{app:stability}
\setcounter{equation}{0}

For the unreduced three-species model \eqref{eq31}, the Jacobian is
\begin{equation}\label{eqA1}
\begin{gathered}
\mathbf J = \begin{pmatrix} J_{11} & -a_{12}x_1 & -a_{13}x_1 \\ e_{12}a_{12}x_2 & J_{22} & -a_{23}x_2 \\ e_{13}a_{13}x_3 & e_{23}a_{23}x_3 & J_{33} \end{pmatrix}, \\ J_{11} = B_1 - 2a_{11}x_1 - a_{12}x_2 - a_{13}x_3, \\
J_{22} = B_2 - 2a_{22}x_2 + e_{12}a_{12}x_1 - a_{23}x_3, \\
J_{33} = B_3 - 2a_{33}x_3 + e_{13}a_{13}x_1 + e_{23}a_{23}x_2,
\end{gathered}
\end{equation}
which reduces to \eqref{eq11} for $N=3$ under the assumptions of Section~\ref{sec:reduced}. Linear stability is a local statement about the linearisation $\delta\dot{\mathbf x}=\mathbf J\,\delta\mathbf x$ at a fixed point. A hyperbolic fixed point in three dimensions is stable if all eigenvalues have negative real part, unstable if all have positive real part, and a saddle otherwise. When the eigenvalues cannot be read off directly, we write the characteristic polynomial of the Jacobian matrix as
\begin{equation}\label{eqA2}
\lambda^3 + c_2\lambda^2 + c_1\lambda + c_0 = 0.
\end{equation}
By the Routh-Hurwitz criterion, all three roots have negative real part if and only if $c_2>0$, $c_0>0$ and $c_2c_1>c_0$. More generally, the number of roots with positive real part equals the number of sign changes in the first column of the Routh array,
\begin{equation}\label{eqA3}
\begin{array}{c|cc} \lambda^3 & 1 & c_1 \\ \lambda^2 & c_2 & c_0 \\ \lambda^1 & (c_2c_1 - c_0)/c_2 & 0 \\ \lambda^0 & c_0 & \end{array}
\end{equation}
so that zero sign changes indicate a stable node or focus, one or two a saddle, and three an unstable node or focus. Finally, the discriminant distinguishes nodes from foci.
\begin{equation}\label{eqA4}
\Delta = 18\,c_2c_1c_0 - 4\,c_2^3c_0 + c_2^2c_1^2 - 4\,c_1^3 - 27\,c_0^2
\end{equation}
$\Delta>0$ gives three distinct real roots, $\Delta<0$ one real root and a complex-conjugate pair, and $\Delta=0$ a repeated root.

\end{appendices}

\section*{Statements and Declarations}





\bmhead{Use of AI tools} During the preparation of this manuscript, the authors used an AI-based language model for language editing, formatting of the manuscript and assistance with writing parts of the plotting and simulation code. After using this tool, the authors reviewed, tested and edited the content and code as needed and take full responsibility for the content of the publication.

\bibliography{Refs}

\end{document}